\documentclass[reprint,aps,superscriptaddress,amsmath,amssymb]{revtex4-2}
\usepackage[latin9]{inputenc}
\usepackage{color}
\usepackage{graphicx}
\usepackage{esint}
\usepackage[linkcolor=blue,urlcolor=blue,citecolor=blue,colorlinks=true]{hyperref}
\usepackage[dvipsnames]{xcolor}
\usepackage[normalem]{ulem}

\def\ESB{E_\textsc{sb}}
\def\EBD{E_\textsc{b}}
\def\PBD{\phi_\textsc{b}}
\def\SBD{\psi_\textsc{b}}
\def\PPG{\phi^p_g}
\def\PPZ{\phi^p_0}
\def\PSP{\Psi^p}

\def\ai{{\it ab initio}}
\def\QF{q_\textsc{F}}
\def\EPOL{E_{\textsc{p}}}
\def\VPOL{\varepsilon_{\textsc{p}}}
\def\VPOD{\dot{\varepsilon}_{\textsc{p}}}
\def\TMAX{T_{\rm max}}
\def\EMAX{E_{\rm max}}
\def\EPSX{\epsilon _{\rm max}}
\def\EMID{E_{\rm mid}}
\def\EPSD{\epsilon _{\rm mid}}
\def\EVAC{E_{\rm vac}}
\def\G1D{G_{\rm 1D}}
\def\KS{\mathbf{k}_\parallel}
\def\KV{\mathbf{k}}
\def\RV{\mathbf{r}}
\def\QV{\mathbf{q}}
\def\ETAR{\eta_{\textsc{r}}}
\def\ETAT{\eta_{\textsc{t}}}
\def\ET1D{\eta_{\textsc{1D}}}
\def\DTAR{\Delta\tau_{\textsc{r}}}
\def\DTAT{\Delta\tau_{\textsc{t}}}
\def\YDIR{\hat{\mathbf y}}
\def\XDIR{\hat{\mathbf x}}
\def\ZDIR{\hat{\mathbf z}}

\begin{document}
\title{Fano physics behind the N-resonance in graphene}
\author{R.O.\ Kuzian}
\affiliation{Donostia International Physics Center (DIPC),
 Paseo Manuel de Lardizabal 4, San Sebasti\'an/Donostia,
  20018 Basque Country, Spain}
\affiliation{Institute for Problems of Materials Science Krzhizhanovskogo 3, 03180
Kiev, Ukraine}
\author{D.V.\ Efremov}
\affiliation{Leibniz Institute for Solid State and Materials Research Dresden, Helmholtzstrasse 20, D-01069 Dresden, Germany}
\author{E.E.\ Krasovskii}
\affiliation{Donostia International Physics Center (DIPC),
 Paseo Manuel de Lardizabal 4, San Sebasti\'an/Donostia,
  20018 Basque Country, Spain}
\affiliation{Departamento de Pol{\'i}meros y Materiales Avanzados: F{\'i}sica, Qu{\'i}mica y
  Tecnolog{\'i}a, Universidad del Pa{\'i}s Vasco-Euskal Herriko Unibertsitatea, Donostia-San
  Sebasti\'an, 20080 Basque Country, Spain}
\affiliation{IKERBASQUE, Basque Foundation for Science, 48013 Bilbao, Spain}

\begin{abstract}
Bound states and scattering resonances in the unoccupied continuum of a
two-dimensional crystal predicted in [Phys.Rev. B {\bf 87}, 041405(R) (2013)]
are considered within an exactly solvable model. A close connection of the observed
  resonances with those arising in the Fano theory is revealed. The resonance occurs when the
  lateral scattering couples the layer-perpendicular incident electron wave to a strictly bound
  state. The coupling strength determines the location of the pole in the scattering
amplitude in the complex energy plane, which is analytically shown to lead
to a characteristic Fano-lineshape of the energy dependence of the electron
transmissivity through the crystal. The implications for the timing of the
resonance scattering are discussed. The analytical results are illustrated
by \ai\ calculations for a graphene monolayer.
\end{abstract}
\date{\today}
\maketitle

\section{Introduction}
The interest in the electronic properties of low-dimensional systems
  has been growing in the last decades owing to the plethora of unique
  quantum phenomena impossible in the three-dimensional (3D) extended
  systems~\cite{Giamarchi2003,Avouris2017,Rocca2020}. A striking example are
  the bound states and scattering resonances in the unoccupied continuum of
  two-dimensional (2D) crystals~\cite{Nazarov2013}. In an infinite 3D crystal
  bound states cannot exist at any energy, whereas in a stand-alone 2D layer
  the electrons below the vacuum level $\EVAC$ are bound to the layer and
  those above $\EVAC$ can escape to infinity. However, as was first shown in
  Ref.~\cite{Nazarov2013} in atomically thin layers solutions of the Schr\"odinger
  equation may exist above the vacuum level that are bound to the layer, thereafter
  referred to as N-states. Such solutions may occur both at complex and at real
  energies, the former giving rise to scattering resonances and the latter being
  true stationary states---a rare example of a bound state in continuum (BIC)~\cite{Hsu2016}.
 In contrast to von Neumann and Wigner BIC
 \cite{vonNeumann1993,Stillinger75,Herrick1976,Stillinger1976} in a 1D continuum and to those
 arising from the interaction of Feshbach's resonances in the scattering on a spherically
 symmetric potential \cite{Fonda1960,Friedrich1985,Friedrich1985b}, the N-states are extended
 along the crystal plane and bound in the perpendicular direction. In
 Ref.~\onlinecite{Nazarov2013}, scattering resonances related to complex-energy N-states were
 predicted to exist in graphene, where they were soon observed
 experimentally~\cite{Pisarra2014,
 Wicki-16,Krivenkov2017,Matta2022,deJong2018,Da2020,Neu2021,Ewert2021,Hong2023,Corso2024} and
further studied
theoretically~\cite{Pisarra2014,Kogan2014,Kogan2017,Ueda2018,Grozdanov2022,Grozdanov2024}.

Conventional band structure or low energy electron diffraction (LEED) calculations are
  performed for real energies, so the parameters of the resonance cannot be directly inferred.
  This calls for the development of an analytical model that would provide a functional relation
  between the complex energy of the resonance and the observables. In the original theory, the
resonances appeared as a result of sophisticated analytical computations for a specific model or
as an \ai\ numerical outcome, and although a Fano-like lineshape of the electron transmission
spectrum was observed in Ref.~\cite{Nazarov2013} both in the analytical model and in graphene,
no explanation for this shape has been suggested and its relation to the properties of the
N-states has not been considered. Here, we fill this gap by considering a simple model
  that elucidates a close connection of the N-resonance with that in the Fano model. We
analytically
show that the transmission amplitude near the resonances has a Fano character~\cite{Fano1961}
\begin{equation}
t(E)  =t_0\frac{E-E_0}{E-\EPOL+i\Gamma/2},\label{eq:tFanoamp}
\end{equation}
where $\EPOL-i\Gamma/2$ is the pole location in the complex energy plane, and $E_0$ is the
point of total reflection, the maximum transmissivity being unity. Real graphene will be shown
to closely follow this formula.

Apart from the transmission probability $T(E)$, an important observable is the
transmission timing. With the recent development of ultrashort laser pulses, the dynamics
of Fano resonances in atoms was studied in real
time~\cite{Ott2014,Deshmukh2018,Banerjee2019,Deshmukh2021EPJ}. It is characterized by
the wave-packet group delay introduced by Bohm, Wigner, Eisenbud, and Smith
\cite{Bohm1951,Wigner55,Smith60} (Wigner time delay), which can be obtained as the energy
derivative of the scattering phase $\eta = \arg(t)$, 
\begin{equation}
\Delta\tau=d\eta/dE\equiv\dot{\eta}.\label{eq:EWS}
\end{equation}
Hartree atomic units $\hbar=e=m_{e}=1$ are used throughout the paper, so $\Delta\tau$ is
measured in the units of $\hbar/1{\rm Ha}=\hbar^3/m_{e}e^4=24.19$~as. Similar to the
atomic case~\cite{Deshmukh2018}, the resonant
part of the delay is a Lorentzian function with the maximum at $\EPOL$ and width $\Gamma$.

The paper is organized as follows. In Sec.~\ref{sec:Model} we introduce a 2D model for
electron scattering on a 1D crystal
(a thin wire) characterized by a sharp attractive potential in the $\YDIR$ direction
and a weak corrugation  along $\XDIR$.
We present the
solution of the model in a general form, which is analyzed
in the following sections. Section~\ref{sec:BIC} discusses the system without
corrugation in terms of eigenstates extended and localized in the $\YDIR$ direction. In
Sec.~\ref{sec:Umklapp}, the corrugation is shown to couple the
extended states with the localized ones 
of the same parity, which leads to Fano scattering resonances.
The timing of the wave packet scattering
is considered in Sec.~\ref{sec:Wigner}. In Sec.~\ref{sec:2wires}, the formalism
is generalized to two identical
wires. Sec.~\ref{sec:Concl} summarizes the results. The details of the 
calculations are given in
Appendices~\ref{sec:Fano-Hamiltonian}--\ref{sec:BS}.

\section{Infinitely thin wire}\label{sec:Model}

To understand the physics of the resonance states~\cite{Nazarov2013},
let us consider an infinite wire along the $\XDIR$ axis described by
a simple 2D Hamiltonian 
\begin{align}
\hat{H} & =\frac{\hat{p}_{x}^2+\hat{p}_{y}^2}{2}+\hat{V} ,\label{eq:HN2D}\\
\hat{V} & = [2\Omega\cos (Kx)-\kappa]\delta(y) \label{eq:V}
\end{align}
An attractive
potential $\kappa>0$ is modeled by a $\delta$-function in the $\YDIR$ direction and a weak
corrugation $\cos(Kx)$ along $\XDIR$, where $K=2\pi/a$, and $a$ is the lattice constant.
In the following we assume that $|\Omega|\ll\kappa$ and consider the case $\kappa<K$,
which somewhat simplifies the formulas. We will show that the continuous spectrum of
the Hamiltonian~(\ref{eq:HN2D}) possesses sharp resonances at positive energies, which makes
it similar to the Fano Hamiltonian $\hat{H}_{F}$ of Ref.~\cite{Fano1961}, see Eq.~(\ref{eq:HF})
in Appendix~\ref{sec:Fano-Hamiltonian}.

We consider the scattering of a normally incident wave, along $\YDIR$.
Similar to our previous work \cite{Krasovskii2024n}, we substitute
the Laue representation for the scattering state 
\begin{align}
\Psi(\RV)=\sum\limits_g\phi_g(y)\exp (igx),\label{eq:Laue} \\
 g=Kn,\quad n=0,\pm1,\pm2,\ldots \nonumber
\end{align}
into both sides of the Lippmann-Schwinger equation 
\begin{widetext}
\begin{align}
\Psi(\RV) &=\exp(ik_yy)+ \delta\Psi(\RV) \label{eq:LipSha} \\   
\delta\Psi(\RV) &= \iint\!\!d\RV'\;G_0\left(\RV-\RV';E\right)\hat{V}(\RV')\Psi(\RV') = 
\int\!\!dx'G_0\left(x-x',y;E\right)\left[2\Omega\cos(Kx')-\kappa\right]\Psi(x',0),
\label{eq:DPsi}
\end{align}
where $E=k_y^2/2$ and $G_0$ is the free-electron Green's function,
\begin{equation} 
G_0\left(\RV;E\right)=\iint\frac{d\QV}{4\pi^2}\frac{\exp(i\QV\RV)}{E-q^2/2}. \label{eq:G02D}
\end{equation}
After the integration
over $x'$ and $\QV$ Eq.~(\ref{eq:LipSha}) reduces to an algebraic equation 
\begin{equation}
\Psi(\RV)  =\sum_g\phi_g(y)\exp(igx)=\exp(ik_yy)\,+ 
  \sum_g\exp(igx)\G1D\left(y,E-g^2/2\right)
  \left[\Omega\left(\phi_{g-K}+\phi_{g+K}\right)-\kappa\phi_g\right],\label{eq:Laue2D}
\end{equation}
\end{widetext}
where $\phi_g\equiv\phi_g(y=0)$ and $\G1D(y,E)$ is the free-electron Green's
function in 1D~\cite{Economou} 
\begin{align}
\G1D(y,E) & =\begin{cases}
-\dfrac{\exp(-k_0|y|)}{k_0}, & E<0,\\[10pt]
-\dfrac{i\exp(ik_0|y|)}{k_0}, & E>0,
\end{cases}\label{eq:g01D}\\
k_0 & \equiv\sqrt{2\left|E\right|}.\label{eq:k0}
\end{align}
It follows from Eq.~(\ref{eq:g01D}) that the $y\rightarrow\pm\infty$ asymptotics of the
scattering function $\Psi(\RV)$ of Eq.~(\ref{eq:Laue2D}) contains only propagating waves
with the positive energetic argument $E-g^2/2$. Thus, in the asymptotic region, the sum
over $g$ contains only finite number of terms with $g^2<2E=k_y^2$. In particular, for
$k_y<K$, i.e., for $E< \ESB =K^2/2$ it contains only the central beam, $g=0$, 
\begin{align}
  \phi_0(y) &=\exp(ik_yy)\, \nonumber\\
  & + \G1D(y,E)
 \left[\Omega\left(\phi_{-K}+\phi_{K}\right)-\kappa\phi_0\right]. \label{eq:cbeam}
\end{align}
Above $\ESB$, secondary beams emerge, with $g>0$.
In other words, the $g>0$ terms in Eq.~(\ref{eq:Laue2D}) become propagating waves.
We equate the coefficients
of the Fourier harmonics $\exp(igx)$ in Eq.~(\ref{eq:Laue2D}) for
$y=0$ and obtain the recurrence relation for the coefficients $\phi_g$:
\begin{align}
\phi_g(1+\kappa F_g) & =\delta_{g,0}+\Omega F_g\left(\phi_{g-K}
+\phi_{g+K}\right),\label{eq:recfig}\\
F_g & \equiv \zeta_g^{-1} \equiv \G1D\left(0,E-g^2/2\right).\label{eq:Fp}
\end{align}
The solution of the infinite three-term recurrence relation~(\ref{eq:recfig})
has the form of a continued fraction \cite{Haydock1980,Viswanath1994}:
\begin{align}
\phi_K & =\cfrac{\Omega\phi_0}{\zeta_K+\kappa-\Omega^2\theta_2}, \label{eq:fiKcfT}\\
\phi_0 & =\cfrac{\zeta_0}{\zeta_0+\kappa-\cfrac{2\Omega^2}{\zeta_K+\kappa-\Omega^2\theta_2}},
\label{eq:fi0cfT}
\end{align}
where $\theta_2$ is the second-step terminator of the continued fraction~(\ref{eq:fi0cfT}),
\begin{equation}
\theta_2 = \cfrac{1}{\zeta_{2K}+\kappa
-\cfrac{\Omega^2}{\zeta_{3K}+\kappa-\cfrac{\Omega^2}{\ddots}}}, \label{eq:th2}
\end{equation}
see the derivation in Appendix~\ref{sec:Calc-of-coefficients}. Due to the symmetry of the
Hamiltonian,
the scattering solution is an even function of $x$, $\Psi(-x,y)=\Psi(x,y)$, so the
coefficients of the Laue representation~(\ref{eq:Laue2D}) satisfy the relation
$\phi_{nK}=\phi_{-nK}$. They can be calculated
with any desired accuracy. By truncating the chain, i.e., by putting $\phi_{n_0K}=0$ at some
$n_0$, we obtain approximate values of $\phi_{nK}$ for $n<n_0$. 

To obtain the functions $F_g$ we substitute Eq.~(\ref{eq:g01D}) into Eq.~(\ref{eq:Fp}).
For propagating $g$-beams $F_g$ are purely imaginary, in particular, for the central
beam it is
\begin{equation}
F_0  =-i/k, \label{eq:F0}
\end{equation}
and for $E<\ESB$ all the other $F_g$ are real,
\begin{equation}
F_{nK} =-1/\sqrt{(nK)^2-k^2}, \quad n>0. \label{eq:FnK}
\end{equation}

The wave function for the central beam is 
\begin{align}
\phi_0(y) & =\exp(ik_yy)\,+r\exp(ik_y|y|)\label{eq:F0xy}\\
 & =\begin{cases}
\exp(ik_yy)+r\exp(-ik_yy), & y\rightarrow-\infty,\\
t\exp(ik_yy), & y\rightarrow+\infty,
\end{cases}
\end{align}
where $r\equiv |r|\exp(i\ETAR)$, $t\equiv |t|\exp(i\ETAT)$ are 
the reflection and transition complex amplitudes. From Eq.~(\ref{eq:cbeam}) we obtain
\begin{equation}
r =F_0\left(2\Omega\phi_{K}-\kappa\phi_0\right),\quad t=1+r.\label{eq:rt}
\end{equation}
The energy dependence of the scattering phase shifts $\ETAR$ and $\ETAT$ determines
the Wigner time delays of the reflected $\DTAR$ and transmitted $\DTAT$
wave packets, see Eq.~(\ref{eq:EWS}).

\section{Bound states above vacuum level}\label{sec:BIC}

Besides the propagating states, which are expressed by
Eq.~(\ref{eq:Laue2D}), the Schr\"odinger equation has solutions bound in the $\YDIR$ direction.
These states can be found from the Lippmann-Schwinger equation (\ref{eq:LipSha})
without the incident wave $\Psi(\RV)=\delta\Psi(\RV)$
with $\delta\Psi(\RV)$ given by Eq.~(\ref{eq:DPsi}).

Let us first neglect the corrugation, i.e., set
$\Omega=0$ in Eq.~(\ref{eq:V}). In this case the recurrence relation
(\ref{eq:recfig}) reduces to $\phi_g(1+\kappa F_g)=0$ and has nontrivial solutions
$\PBD\neq 0$ for energies $\EBD(g)=(g^2-\kappa ^2)/2$ at which $1+\kappa F_g=0$, so
\begin{equation}
\PBD(y) = -\kappa \G1D\left(y,\EBD(g)-\frac{g^2}{2}\right)\PBD
=\sqrt{\kappa}e^{-\kappa\left|y\right|}, \label{eq:1DPBD}
\end{equation}
where $\sqrt{\kappa}$ is a normalization coefficient. The wave function $\PBD(y)$
does not depend on $g$ because $\EBD(g)-g^2/2=-\kappa ^2/2$. For $g^2 > \kappa^2$
these states have positive energies and are bound in $\YDIR$ direction. Indeed, for
$\Omega=0$ the motions along $\XDIR$ and $\YDIR$ are independent. The wave functions
are a product of the plane waves $\exp(ik_xx)$ propagating along $\XDIR$ and the
eigenstates of the 1D Hamiltonian of the motion along $\YDIR$
\begin{equation}
\hat{H}_y=\frac{\hat{p}_{y}^2}{2}-\kappa\delta(y).\label{eq:H0y}
\end{equation}
The eigenstates include one bound state $\PBD(y)$ at the energy
$\EBD=-\kappa^2/2$. This yields the eigenstates of the full 2D Hamiltonian
$\hat{H}=\hat{p}_{x}^2/2+\hat{H}_y$ that are bound in the $\YDIR$ direction
and propagate along $\XDIR$, 
\begin{equation}
\SBD(k_x,\RV)=\exp(ik_xx)\PBD(y).\label{eq:psiqb}
\end{equation}
For a particular case of $k_x=0$ the bound states with energies
$\EBD(g)=(g^2-\kappa ^2)/2$ have been obtained above from the 
Lippmann-Schwinger equation. For a general $k_x$ their energy is
$\EBD(k_x)=(k_x^2-\kappa^2)/2$.

Apart from the bound state $\PBD$, see Eq.~(\ref{eq:1DPBD}), the 1D
Hamiltonian $\hat{H}_{y}$ has the continuum 
eigenstates with positive energies $E(k_y)=k_y^2/2$, which may be odd
$\varphi^-(y)=\sqrt{2}\sin(k_yy)$ or even
$\varphi^+(y)=\sqrt{2}\cos(k_y|y|+\ET1D)$
under the reflection $y\to-y$. The
phase shift $\ET1D$ is given by the equations 
\begin{equation}
\cos(\ET1D) = \frac{k_y}{\sqrt{k_y^2+\kappa^2}},\quad
\sin(\ET1D) = \frac{\kappa}{\sqrt{k_y^2+\kappa^2}}.\label{eq:etak}
\end{equation}
The functions $\varphi^-$ and $\varphi^+$ multiplied by plane waves are then
eigenstates of the full 2D Hamiltonian $\hat{p}_{x}^2/2+\hat{H}_y$ with positive
energies $E(\KV)=k^2/2$,
\begin{equation}
\psi^p(\KV,\RV)  =\exp(ik_xx)\varphi^p(k_y,y), \label{eq:psiqks}
\end{equation}
where $p=\pm$ is the parity index.

\begin{figure*}[htb] %%%%%%%%%%%%%%%%%%%%%%%%%%%%%%%%%%%%%%%%%%%%%%%%%%%%%%%%%%%%%%%% 1
\includegraphics[width=0.9\columnwidth]{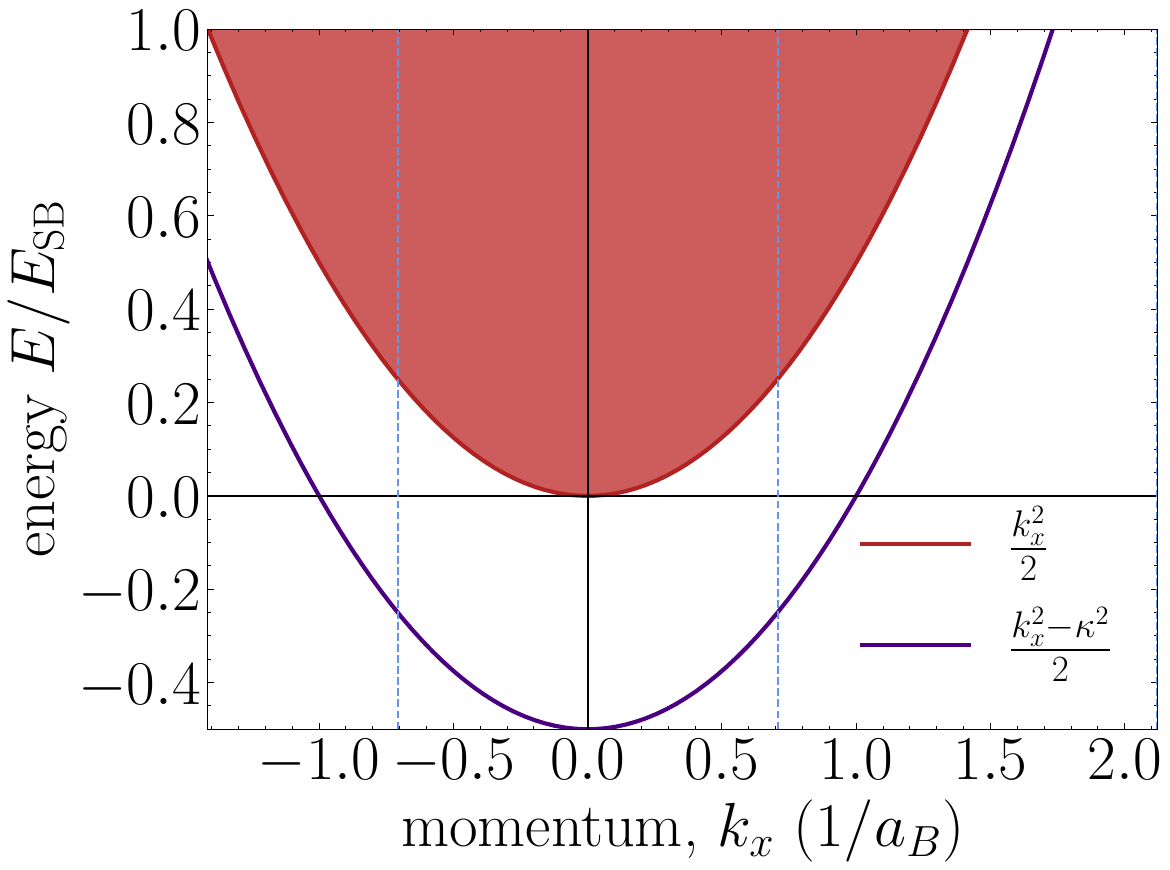} \hspace{10mm} 
\includegraphics[width=0.9\columnwidth]{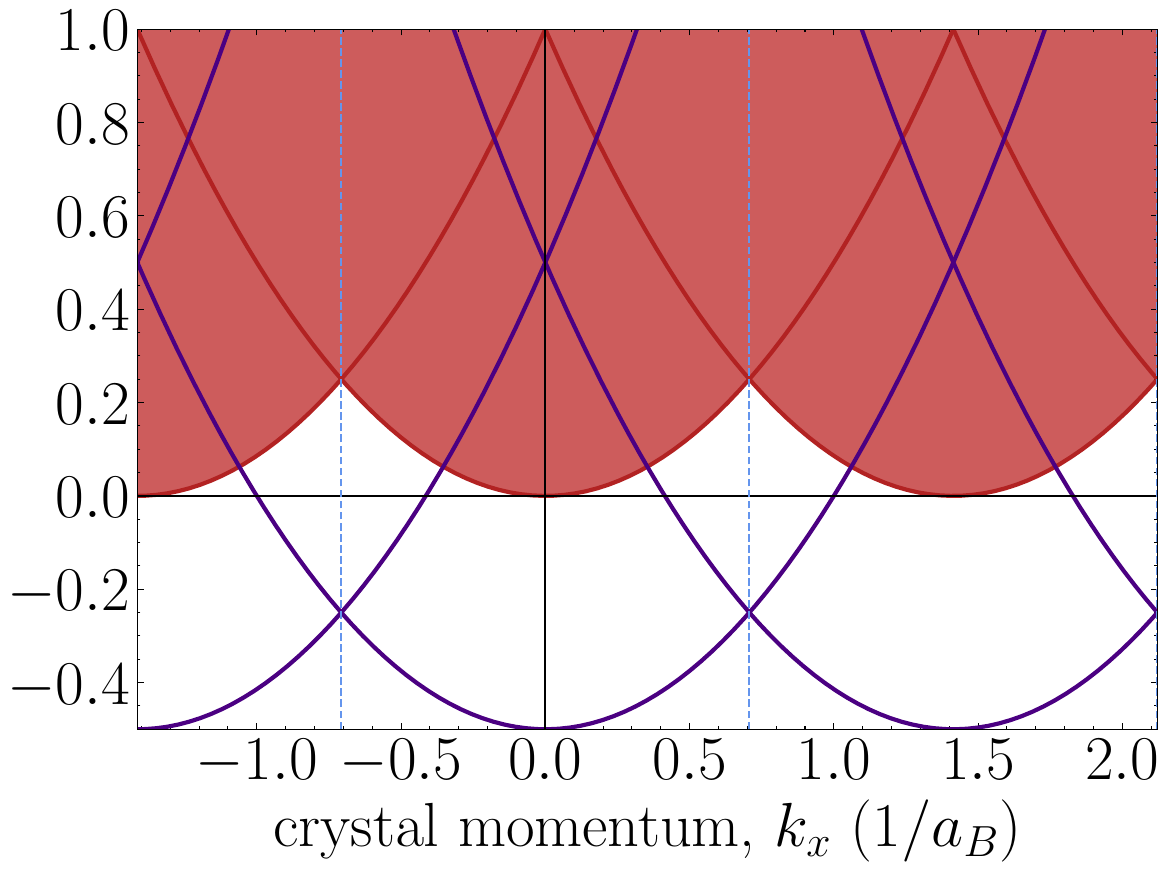} 
\caption{\label{f:spH0} Band structure of the model in the extended (left panel) and
  reduced (right panel) zone schemes in the absence of corrugation, $\Omega=0$. Other
  parameters of the model are $K=\sqrt{2}/a_B$, $\kappa=1/a_B$.}
\end{figure*} %%%%%%%%%%%%%%%%%%%%%%%%%%%%%%%%%%%%%%%%%%%%%%%%%%%%%%%%%%%%%%%%%%%%%%% 1
In the left panel of Fig.~\ref{f:spH0}, the band structure is depicted in 
the extended zone scheme. The right panel of Fig.~\ref{f:spH0} shows the same plot in the 
reduced zone scheme for an artificial periodicity in the $\XDIR$ direction. 
For a given crystal momentum $k_{x}$ in the first Brillouin zone
the spectrum contains discrete bound states    
$\SBD(k_n,\RV)$, $k_n=k_{x}+Kn$ and continuum of even 
$\psi^+(k_n,k_y,\RV)$ and odd $\psi^-(k_n,k_y,\RV)$ extended states with
continuously varying $k_y$.

The states $\psi^+(\KV,\RV)$ and $\psi^-(\KV,\RV)$ 
may be combined to form scattering states. In the absence of the corrugation,
the normally incident wave produces only the specularly reflected beam, and
the scattering states have the form (\ref{eq:F0xy}). For $\Omega=0$, the chain
of equations (\ref{eq:recfig}) contains only the first equation with $g=0$, 
\[
\phi_0\bigg|_{\Omega=0}=\frac{1}{1+\kappa F_0}.
\]
Equation~(\ref{eq:rt}) gives the transmission coefficient 
\begin{equation}
t_0=\frac{1}{1+\kappa F_0}=\frac{k_y}{k_y-i\kappa}
=\cos(\ET1D)e^{i\ET1D},\label{eq:t0}
\end{equation}
which is a smooth function of $k_y$ and $E$.

The right panel of Fig.~\ref{f:spH0} shows that in the repeated zone scheme,
for $\EBD(k_n)=E(\KV)$, i.e., $k_n^2-\kappa^2 =k^2$ the states $\SBD(k_n,\RV)$ are
degenerate with the states $\psi^p(\KV,\RV)$. As we will see in the next section,
the corrugation couples the extended and the even bound state and transforms it
into a Fano resonance, whereas the odd state remains bound in the $\YDIR$ direction.
In this case the transmission probability $T=|t|^2$ acquires a
characteristic Fano shape near the resonance, in contrast to the smooth behavior
of $T_0=|t_0|^2$.

\section{Coupling by Umklapp processes\label{sec:Umklapp}}

The corrugation $\hat{\Omega}=2\Omega\cos(Kx)\delta (y)$ couples the
bound states $\SBD(k_x,\RV)$ of Eq.~(\ref{eq:psiqb}) with the even extended states
$\psi^+(\KV',\RV)$, similar to the Fano Hamiltonian~(\ref{eq:HF}) where the perturbation
$\hat{V}_{F}$ couples a discrete level with the continuum~\cite{Fano1961}:
\begin{align}
& \iint[\psi^+(\KV',\RV)]^*
\hat{\Omega}\SBD(k_{x},\RV)d^2\RV \nonumber \\
& =\omega_{kk'}\left[\delta(k_{x}+K-k_x')
  +\delta(k_{x}-K-k_x')\right], \label{Vcoupled}
\end{align}
where $\omega_{kk'}=2\pi\Omega\sqrt{2\kappa}\cos\ET1D$.
However, our case is more involved because $\hat{\Omega}$ also has nonzero matrix elements
between the states of the same kind.

\begin{figure}[b] %%%%%%%%%%%%%%%%%%%%%%%%%%%%%%%%%%%%%%%%%%%%%%%%%%%%%%%%%%%%%% 2
\centering \includegraphics[width=0.98\columnwidth]{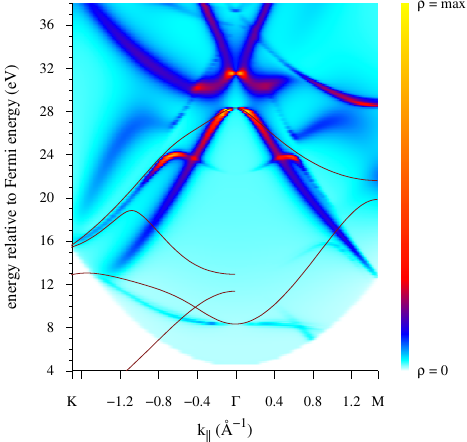} 
\caption{
  Local spectral function $\rho(E,\KS)$ of the LEED states for graphene along
  the $\Gamma M$ and $\Gamma K$ lines. $\rho(E,\KS)$ characterizes the
  dwell time at the scatterer, which peaks at the resonances. The lines superimposed
  on the map show the dispersion of the strictly bound states, which are odd under
  reflection in the plane containing the $\KS$ vector and the surface normal.
}
\label{f:sdmap} 
\end{figure}  %%%%%%%%%%%%%%%%%%%%%%%%%%%%%%%%%%%%%%%%%%%%%%%%%%%%%%%%%%%%%%%%% 2
Let us now consider a normally incident wave along $\YDIR$, i.e., $k_x'=0$ and
$E(\KV')=k_y'^2/2$. Then $\hat{\Omega}$ couples the continuous states $\psi^+(0,k_y',\RV)$
and the bound states $\SBD(\pm K,\RV)$ having the energy $\EBD(\pm K)=(K^2-\kappa^2)/2$. The
resonance occurs near $E_{r}=k_{r}^2/2=\EBD(\pm K)$, where $k_{r}^2=K^2-\kappa^2$. Below, we will
consider the simplest case of $\kappa<K$ and $E_{r}<\ESB$, i.e., the resonance 
in the energy range where only the central beam is present. The generalization to
resonances at higher energies is straightforward.

In a real 3D crystal, the resonances manifest themselves as special features of the Bloch
spectral function $A(E,\KS)$, see Fig.~1 of Ref.~\cite{Nazarov2013}. However, for a strictly
2D crystal the $A(E,\KS)$ function depends on the artificial supercell lattice constant in
the $\ZDIR$ direction. Thus, it is instructive to visualize the resonances in terms of the  
scattering states---the wave functions that describe the LEED experiment. For an
electron of energy $E$ and layer-parallel Bloch vector $\KS$ incident from $z=-\infty$ the
color map in Fig.~\ref{f:sdmap} shows the probability $\rho(E,\KS)$ to find the electron
between the planes $z=-1$ and 4~a.u. embracing the graphene layer located at $z=0$ 
(for the $\rho(z,E)$ function at $\KS=0$ see Fig.~4(b) in Ref.~\cite{Nazarov2013}).
The dispersion of the resonances is distinctly visible as the lines of enhanced dwell time
at the scatterer.

%%%%%%%%%%%%%%%%%%%%%%%%%%%%%%%%%%%%%%%%%%%%%%%%%%%%%%%%%%%%%%%% 3
\begin{figure*}[htb]
\includegraphics[width=0.95\textwidth]{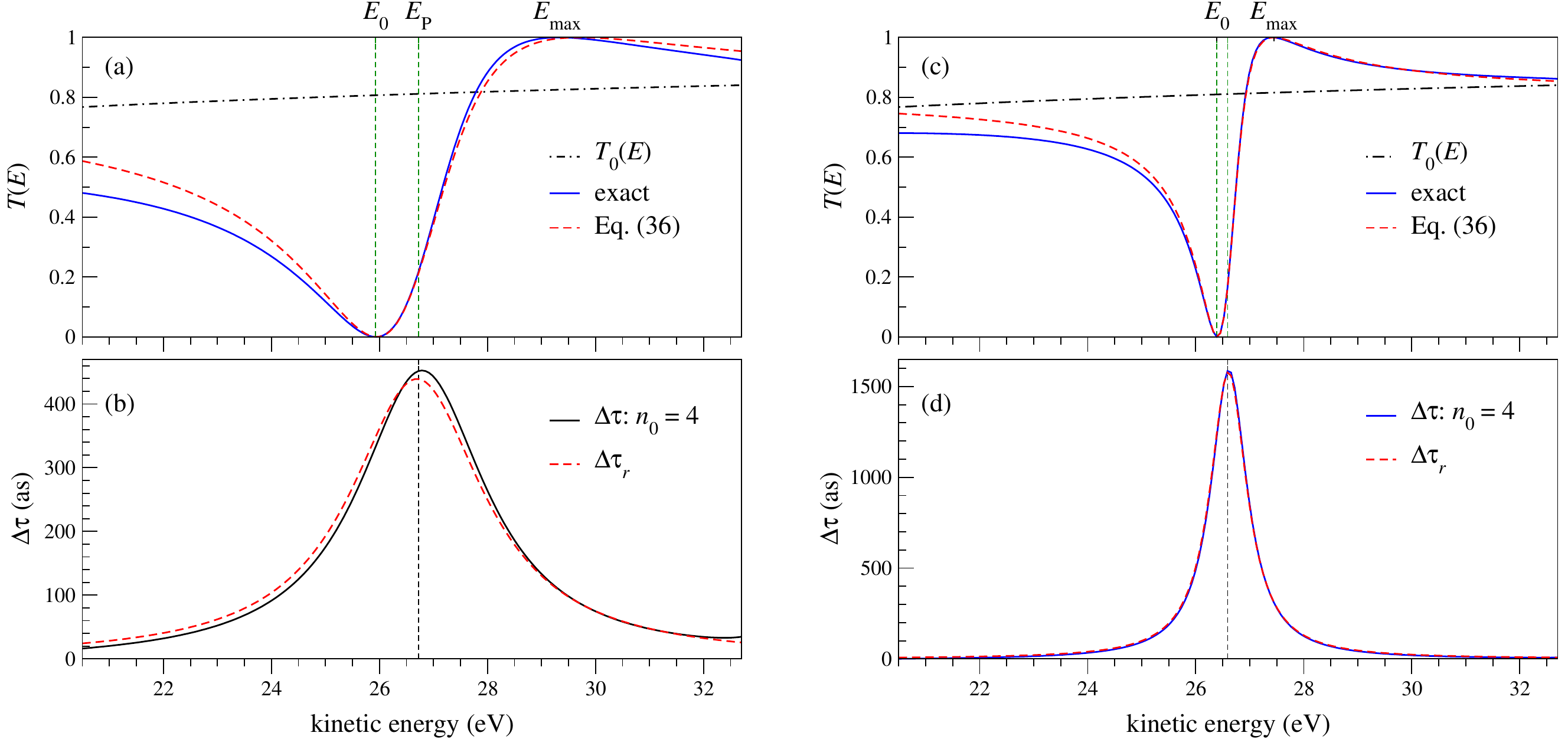} 
\caption{\label{f:tr} (a)~and (c)~Energy dependence of the transmissivity for the
    Hamiltonian~(\ref{eq:HN2D}) with the corrugation parameters $\Omega/\kappa = 0.38$~(a) and
    0.2~(c), see Eq.~(\ref{eq:tfull}). The structureless dash-dotted curve is the transmissivity
    in the absence of the corrugation, $T_0(E)=2E/(2E+\kappa^2)$. 
    Solid lines come from a numerically exact calculation of $\theta_2$ in Eq.~(\ref{eq:th2}).
    The truncation of the chain~(\ref{eq:recfig}) at $n_0=3$ and 4 yields visually
    indistinguishable curves. Dashed curves are the approximation by Eq.~(\ref{eq:tF}).
    (b)~and (d)~Wigner time delay $\DTAT$ (solid line) and its resonant part $\Delta\tau_{r}$
    (dashed line), see Eq.~(\ref{eq:dtaurapp}). Vertical dashed line shows the real part of
    the energy pole $\EPOL$, Eq.~(\ref{eq:Ep}). The model parameters are the same as in
    graphs~(a) and~(c), respectively.}
\end{figure*} %%%%%%%%%%%%%%%%%%%%%%%%%%%%%%%%%%%%%%%%%%%%%%%%%%%%%%%%%%%%%%%% 3
Note that for the Hamiltonian~(\ref{eq:HN2D}) with a finite corrugation
there exist states odd under the reflection $x\to-x$, which
for $E<\ESB$ are necessarily  bound in the $\YDIR$ direction. Indeed, for $k_x'=0$ the
function $\psi^+(0,k_y,\RV)=\varphi^+(k_y,y)$ [see Eq.~(\ref{eq:psiqks})] is obviously
an even function of $x$. Thus, as follows from Eq.~(\ref{Vcoupled}), the corrugation couples
$\psi^+$ only with the even combination $\SBD(K,\RV)+\SBD(-K,\RV)=2\cos(Kx)\PBD(y)$ of the
bound states, whereas their odd combination $\SBD(K,\RV)-\SBD(-K,\RV)=2i\sin(Kx)\PBD(y)$
remains unperturbed and confined in the $\YDIR$ direction. The bound states with positive
energy are a quite general phenomenon, beyond the present model. In particular, they exist
in graphene, see Fig.~\ref{f:sdmap}.

Let us now calculate the transmission amplitude $t$ using the  approximate solution
of the recurrence (\ref{eq:recfig}). Substituting $\phi_K$ and $\phi_0$ from
Eqs.~(\ref{eq:fiKcfT}) and~(\ref{eq:fi0cfT}), respectively,
into Eq.~(\ref{eq:rt}) we obtain after some algebra
\begin{equation}
  t =t_0(k_y)\dfrac{\zeta_K+\kappa-\Omega^2\theta_2}
  {\zeta_K+\kappa-\Omega^2\theta_2-2\Omega^2F_0/(1+\kappa F_0)}, \nonumber
\end{equation}
and substituting $F_0$ from Eq.~(\ref{eq:F0}) we finally obtain
\begin{equation} 
  t =t_0(k_y)\dfrac{\zeta_K+\kappa-\Omega^2\theta_2}
  {\zeta_K+\kappa-\Omega^2\theta_2-\dfrac{2\Omega^2\kappa}{k^2+\kappa^2}+
    \dfrac{2i\Omega^2k}{k^2+\kappa^2}}. \label{eq:tfull}
\end{equation}
In the absence of the corrugation, $\Omega=0$, the amplitude $t=t_0(k_y)$
is a smooth function of energy, see Eq.~(\ref{eq:t0}), and the corrugation
drastically changes its  behavior: The amplitude $t$ acquires a Fano character
near the resonance energy $E_{r}=k_{r}^2/2$, which we now demonstrate based on
Eq.~(\ref{eq:tfull}). Let us consider the energy range $k<K$ where $F_{nK}$
are real for all $n>0$, see Eq.~(\ref{eq:FnK}). For simplicity, we consider
a sufficiently weak potential, $\kappa < K$. At the resonance, the sum
$\zeta_K+\kappa$ goes through zero, so in its vicinity, $E-E_{r} \ll \kappa$,
the Tailor expansion holds:
\begin{equation}
\zeta_K+\kappa=
\kappa-\sqrt{\kappa^2-2(E-E_{r})} \approx (E-E_{r})/\kappa,
\label{eq:FKm1app}
\end{equation}
which immediately yields 
\begin{equation}
t \approx t_0(k_{r})\frac{E-E_0}{E-\EPOL+i(\Gamma/2)}. \label{eq:tF}
\end{equation}
This approximate expression for the transmission amplitude has the canonical
form~(\ref{eq:tFanoamp}) of the Fano resonance~\cite{Fano1961} with the
parameters
\begin{align}
E_0 & = E_r+\kappa \Omega^2\theta_2(E_r),\label{eq:E0}\\
\EPOL & = E_0+\dfrac{2\Omega^2\kappa^2}{k_{r}^2+\kappa^2}=
E_0+\dfrac{2\Omega^2\kappa^2}{K^2},\label{eq:Ep}\\
\frac{\Gamma}{2} & = \frac{2\kappa k_{r}\Omega^2}{k_{r}^2+\kappa^2}=
\frac{2\kappa k_{r}\Omega^2}{K^2}. \label{eq:05Gam}
\end{align}
Note that $\theta_2$ is a smooth function of energy, see Eq.~(\ref{eq:th2}), so 
it can approximated by its value at $E_r$.

\begin{figure} %%%%%%%%%%%%%%%%%%%%%%%%%%%%%%%%%%%%%%%%%%%%%%%%%%%%%%%%%%%%%% 4
\centering \includegraphics[width=0.9\columnwidth]{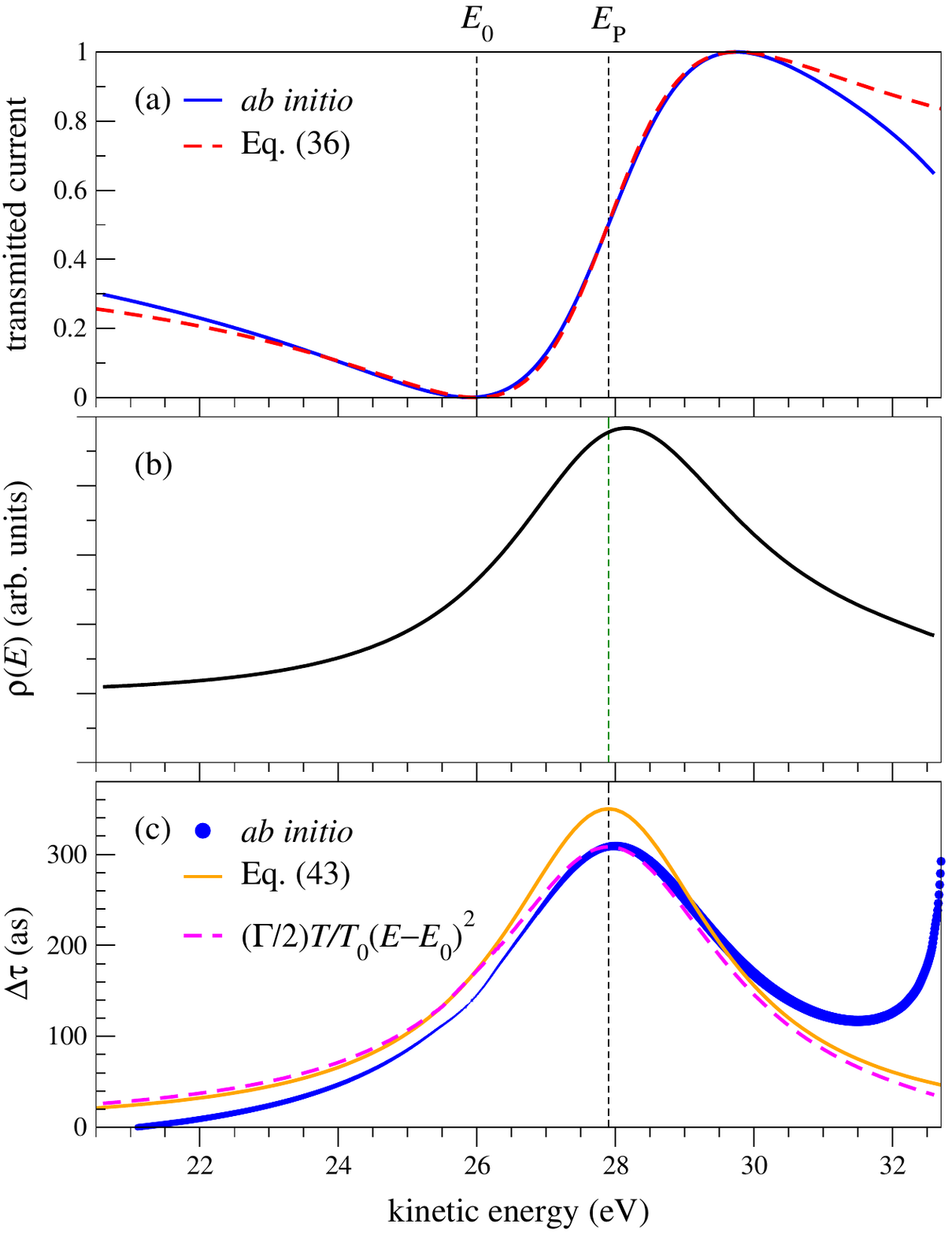} 
\caption{(a)~Transmissivity $T(E)$ of graphene (solid line) and the Fano function of
  Eq.~(\ref{eq:TFmodel}) (dashed line) for the model parameters $E_0=26$~eV,
  $\Gamma=3.76$~eV, and $\EPOL=27.9$~eV, see Appendix~\ref{sec:expFano}.
  (b)~Wigner time delay $\DTAT$ for scattering on graphene [solid line of variable thickness
  proportional to $T(E)$] and delay $\Delta\tau_{r}$ obtained from the Fano function with the
  same parameters as in graph~(a) (thin line). Dashed line shows $\Delta\tau$ derived from
  the \ai\ $T(E)$ curve according to Eqs.~(\ref{eq:TFmodel})
  and~(\ref{eq:dtaurapp}): $\Delta\tau=(\Gamma/2)T(E)/T_0(E-E_0)^2$ assuming $T_0=0.5$.
  Vertical dashed line
  shows the real part of the energy pole $\EPOL$, see Eq.~(\ref{eq:Ep}). (c)~Spectral
  function $\rho(E,\KS)$ of the scattering states for $\KS=0$, see Fig.~\ref{f:sdmap} and its
  caption.}
\label{f:FanoTgr} 
\end{figure}

Near the resonance, the transmissivity is
\begin{equation}
T=|t|^2=T_0(k_{r})\frac{(E-E_0)^2}{(E-\EPOL)^2+(\Gamma/2)^2},\label{eq:TFmodel}
\end{equation}
where $T_0(k_y)=k_y^2/(k_y^2+\kappa^2)$ is the transmissivity in the absence of the
corrugation, according to Eq.~(\ref{eq:t0}). This allows us to express the relation
between the parameters in terms of the transmissivity:
\begin{align}
\EPOL - E_0 & = \dfrac{2\Omega^2\kappa^2}{k_r^2}T_0(k_r),\label{eq:EpT0}\\
\frac{\Gamma}{2} & =\frac{2\kappa \Omega^2}{k_r}T_0(k_r).\label{eq:GamT0}
\end{align}
The maximum of the Fano curve is related to the Fano lineshape asymmetry
\begin{equation}
\QF \equiv(\EPOL-E_0)/(\Gamma/2), \label{eq:qF}
\end{equation}
as $\TMAX =T_0(\QF^2+1)$, see Appendix~\ref{sec:expFano}. From the
expressions~(\ref{eq:EpT0}) and (\ref{eq:GamT0}) for $\EPOL - E_0$
and $\Gamma$, respectively, we obtain $\QF=\kappa/k_{r}$, so
\begin{equation}
\TMAX =T_0(k_{r})\left[\left(\frac{\kappa}{k_{r}}\right)^2+1\right]=1.\label{eq:Tmax1}
\end{equation}
which proves that it is a point of complete transparency. 
The maximum is reached at $\EMAX = \EPOL+\Gamma/2\QF$, 
and the distance between the points of total reflectivity and complete transparency
is $\EMAX - E_0 = 2\Omega^2$. Figure~\ref{f:tr}(a) shows the transmissivity for
$|\EBD|=\kappa ^2/2= 6.54$~eV, which brings the energy $E_0$ close to its location in
graphene, see Fig.~\ref{f:FanoTgr}(a). To reproduce the width of the resonance in
graphene we must set $\Omega^2=1.88$~eV, Fig.~\ref{f:tr}(a). This value is too large
to be considered a small perturbation, $\Omega/\kappa =0.38$: the
  $T(E)$ maximum $\EMAX=30.06$~eV occurs rather far from $E_r= 26.91$~eV, and the
  approximation (\ref{eq:FKm1app}) becomes less accurate around $\EMAX$. As a result, the
  curve given by Eq.~(\ref{eq:TFmodel}) visibly deviates from the exact $T(E)$. For a weaker
  corrugation, $\Omega = 0.2\kappa $, the two curves coincide over an energy interval
  much wider than $\EMAX-E_0$, see Fig.~\ref{f:tr}(c).

   %%%%%%%%%%%%%%%%%%%%%%%%%%%%%%%%%%%%%%%%%%%%%%%%%%%%%%%%%%%%%% 4
Figure~\ref{f:FanoTgr}(a) compares the \ai\ transmissivity $T$ 
for the graphene monolayer~\cite{Nazarov2013,Krasovskii2024n} with the Fano function. The
Fano parameters $E_0=26.00$~eV, $\EPOL=27.90$~eV, and $\Gamma=3.76$~eV are calculated from the
energies of the minimum of the \ai\ curve $E_0=26.00$~eV, its maximum
$\EMAX=29.73$~eV, and the midpoint $\EMID=27.89$~eV, where $T$ reaches the half-maximum,
as explained in Appendix~\ref{sec:expFano}, in particular see Eq.~(\ref{eq:Dm2Gq}) for
$\Gamma$ and Eq.~(\ref{eq:qG2Ep}) for $\EPOL$. As in the model, in the \ai\
curve the maximum is the point of complete transparency.

\section{Wigner time delay}\label{sec:Wigner}

The Wigner time delay $\Delta\tau$ is defined as the difference in
the time of arrival of a free particle and a scattered one in a region
far from the scatterer. For a spectrally narrow wave packet the delay
equals the energy derivative of the scattering phase \cite{Bohm1951,Wigner55,Smith60},
see Eq.~(\ref{eq:EWS}). In our model, the transmission amplitude is the product
of the amplitude $t_0$ in the absence of the corrugation and the resonant 
part due to the corrugation-induced coupling, see Eqs.~(\ref{eq:t0}), (\ref{Vcoupled}),
and (\ref{eq:tfull}). Thus, the phase $\ETAT=\arg(t)$ is the sum 
\begin{align}
\ETAT & =\ET1D+\eta_{r},\nonumber \\
\tan\eta_{r} & =-\frac{2\Omega^2k}{(k^2+\kappa^2)\left[\zeta_K
+\kappa-\Omega^2\theta_2\right]-2\Omega^2\kappa}\label{eq:tgetar}\\
 & \approx-\frac{\Gamma}{2(E-\EPOL)}.\label{eq:tgetarapp}
\end{align}
The approximate equality~(\ref{eq:tgetarapp}) follows from Eq.~(\ref{eq:tF})
and is valid in the vicinity of the resonance. Hence, the delay is
$\DTAT=\Delta\tau_0+\Delta\tau_{r}$, and the resonant part $\Delta\tau_{r}=\dot{\eta_{r}}$
has a simple Lorentzian form near $\EPOL$:
\begin{equation}
\dot{\eta_{r}}\approx\frac{\Gamma}{2\left[(E-\EPOL)^2
+(\Gamma/2)^2\right]}. \label{eq:dtaurapp}
\end{equation}
The exact general expression for $\Delta\tau_{r}$ for $E<K^2/2$
is presented in Appendix~\ref{sec:delay}. 

The energy dependence of the time delay for our 2D model is shown in Figs.~\ref{f:tr}(b)
and~\ref{f:tr}(d). This model turns out to rather accurately describe the time delay of the wave
packet transmitted through graphene, as demonstrated in Fig.~\ref{f:FanoTgr}(b). 
Using the parameters of the Fano function extracted from the \ai\ $T(E)$ curve of
Fig.~\ref{f:FanoTgr}(a) we obtain quantitative agreement between the \ai\ and the
model $\Delta\tau(E)$ curves. The difference between them is the non-resonant
contribution $\Delta\tau_0$. The maxima of both curves are at $\EPOL\approx27.9$~eV,
which is the real part of the amplitude pole energy. Moreover, the maximum of the local
spectral function $\rho(E,\KS=0)$ is very close to $\EPOL$, and its width is close to that
of the $\Delta\tau(E)$ peak, see Fig.~\ref{f:FanoTgr}(c).

\section{Two parallel wires}\label{sec:2wires}

%%%%%%%%%%%%%%%%%%%%%%%%%%%%%%%%%%%%%%%%%%%%%%%%%%%%%%%%%%%%%%%%
\begin{figure}[b]  %%%% [htb]
\includegraphics[width=0.9\columnwidth]{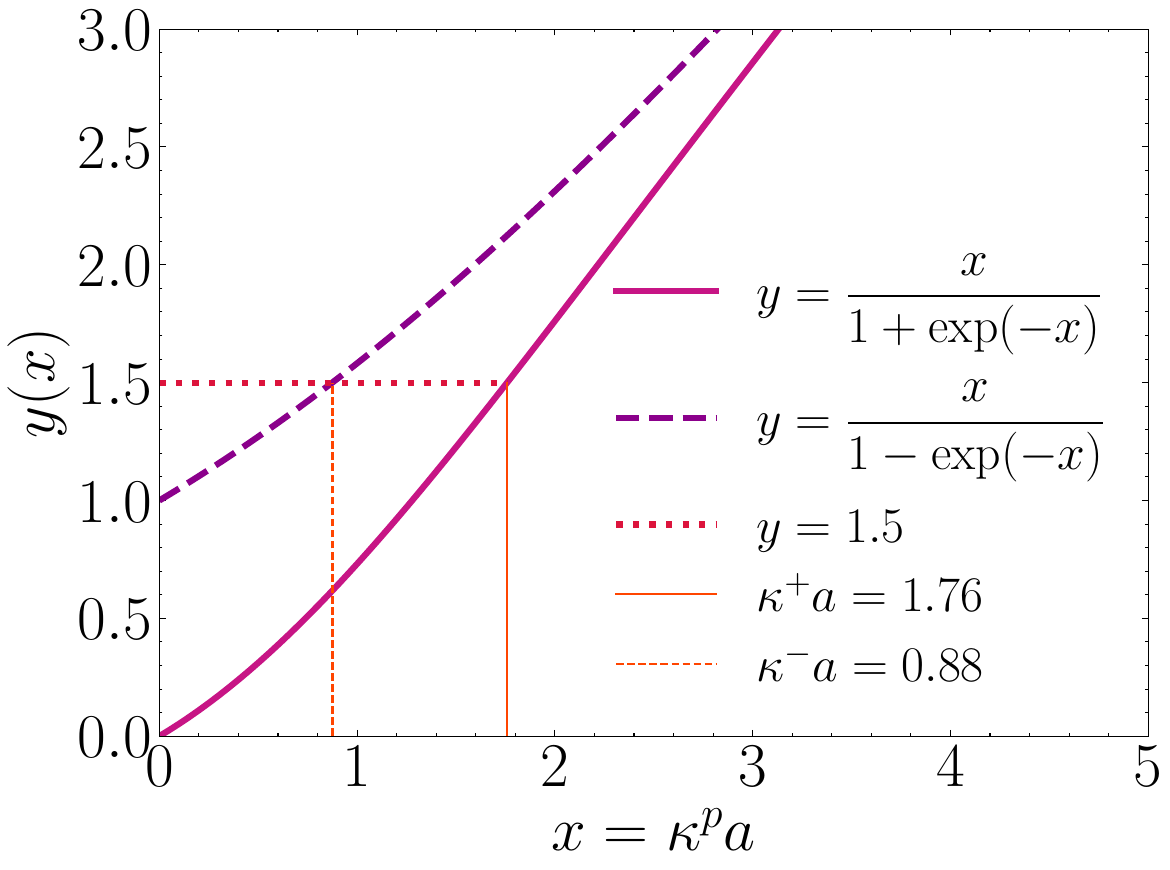}
\caption{\label{f:eb2} The bound state energy  
  parameter $\kappa^p=\sqrt{2|\EBD^p|}$ is determined from the intersection of the 
  curve $y(x)=x/[1+p\exp(-x)]$ with the horizontal line $y(x)=\kappa a$. Odd bound
  states, $p=-$, exists only for $\kappa a > 1$. 
}
\end{figure}
%%%%%%%%%%%%%%%%%%%%%%%%%%%%%%%%%%%%%%%%%%%%%%%%%%%%%%%%%%%%%%%% 
%%%%%%%%%%%%%%%%%%%%%%%%%%%%%%%%%%%%%%%%%%%%%%%%%%%%%%%%%%%%%%%%
\begin{figure*}[htb]
\includegraphics[width=0.9\columnwidth]{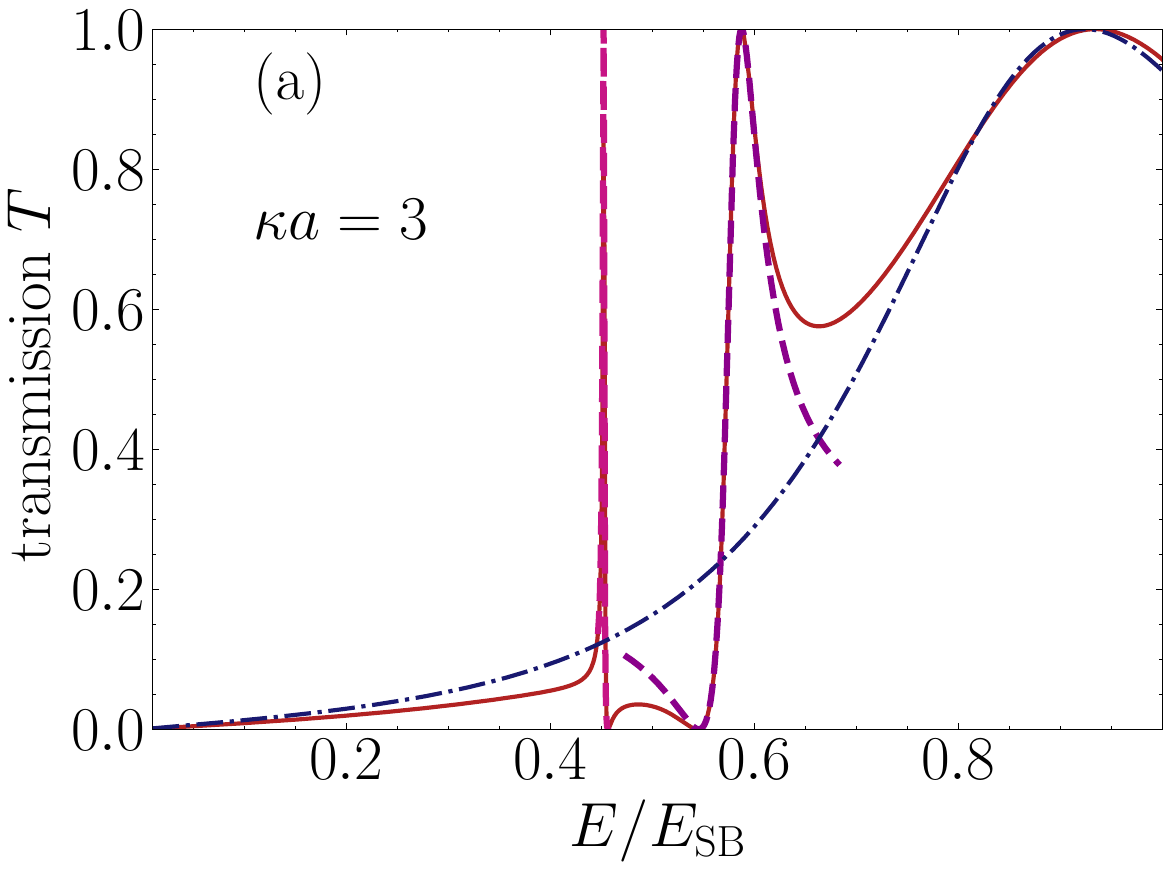} \hspace{10mm}
\includegraphics[width=0.9\columnwidth]{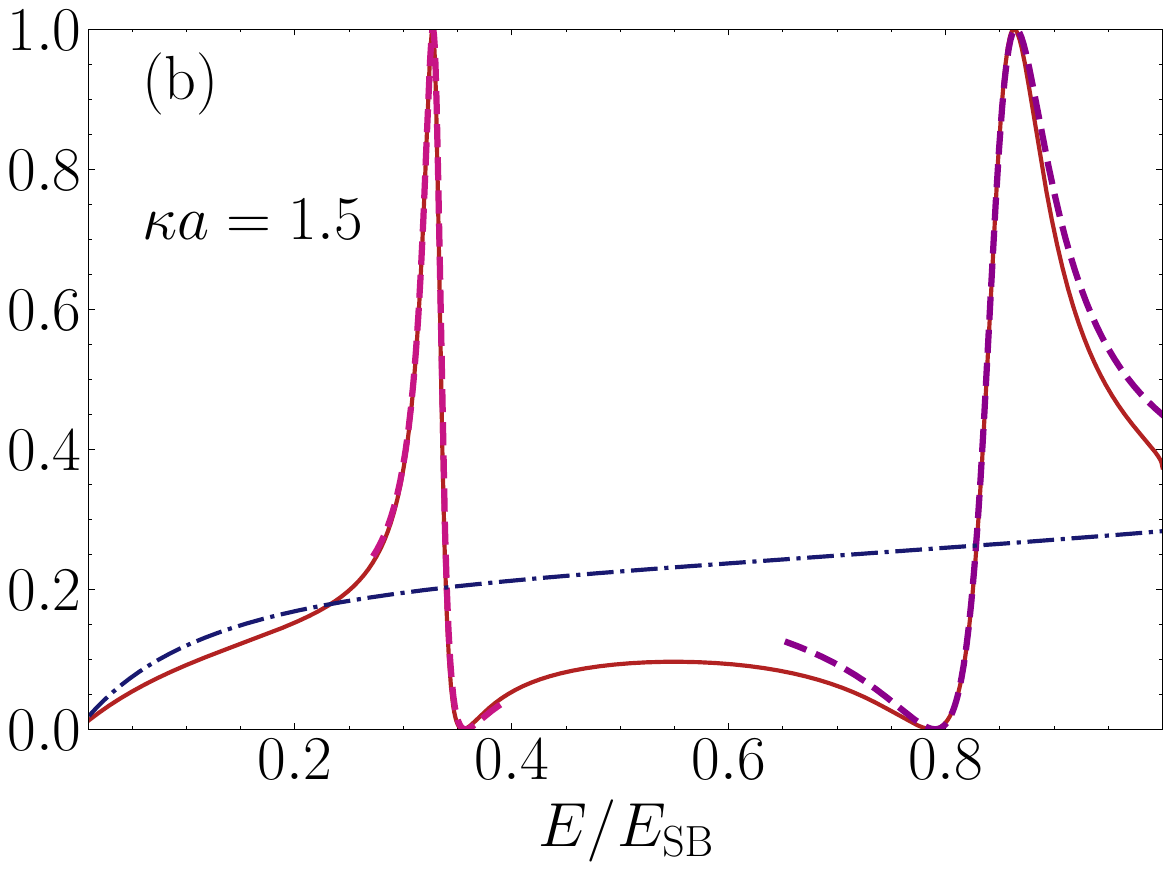} 
\caption{\label{f:t2ls} Energy dependence of the transmissivity $T=|t|^2$ by
  Eq.~(\ref{eq:t2}) for the model Hamiltonian $\hat{H}_2$ of Eq.~(\ref{eq:HN2ch}). Solid line
  shows a numerically exact calculation of $\theta_2^p$. The truncation of the
  chain~(\ref{eq:recfig}) at $n_0=3$ yields a visually indistinguishable curve.
  The blue dash-dotted curve is the transmissivity for the absence of corrugation. The
  approximate $T(E)$ in the vicinity of resonance by Eq.~(\ref{eq:fitildF}) are shown
  by the dashed curves.
}
\end{figure*}
%%%%%%%%%%%%%%%%%%%%%%%%%%%%%%%%%%%%%%%%%%%%%%%%%%%%%%%%%%%%%%%% 
%
For multiple wires, the interference between waves transmitted and reflected from
individual scatterers drastically modifies the energy dependence of the transmission
probability and the time delay~\cite{Krasovskii2024n}. In this case, Eq.~(\ref{eq:Laue2D})
is replaced with a system of equations, generally one equation per wire. Here, we
consider two identical wires and make use of the symmetry of the Hamiltonian,
whereby the two equations decouple. The Hamiltonian for two wires reads
\begin{align}
& \hat{H}_2 =\frac{\hat{p}_{x}^2+\hat{p}_{y}^2}{2}+\hat{V}_2,\label{eq:HN2ch}\\
& \hat{V}_2 =\left[2\Omega\cos(Kx)-\kappa\right]\left[\delta\left(y-\frac{a}{2}\right)+
\delta\left(y+\frac{a}{2}\right)\right], \label{eq:H2}
\end{align}
and we again assume that $\left|\Omega\right|\ll\kappa$.

In the absence of the corrugation, two states bound within the structure
appear, so two resonances could be expected. (Their energies and wave functions
are calculated in Appendix~\ref{sec:BS}). Due to the symmetry of the Hamiltonian,
its eigenfunctions may be chosen to be even or odd functions of both coordinates.
For the plane wave incident along $\YDIR$ it is sufficient to consider
only the eigenfunctions that are even under the reflection $x\to-x$
and may be even or odd in the $y$ variable, $\Psi^p(x,-y) = p \Psi^p(x,y)$. 
The subspaces of even and odd functions are orthogonal, and the Lippmann-Schwinger
equation may be solved for functions in each subspace separately.
The coefficients of Laue representations for the scattering states
acquire additional index $p$ for even $p=+$ and odd $p=-$ functions,
\begin{equation}
\PSP(\RV)=\sum\limits_g\PPG(y)\exp(igx).\label{eq:Laue2}
\end{equation}
The Lippmann-Schwinger equation now has the form 
\begin{widetext}
\begin{align}
\PSP(\RV) & =\PSP_0(y)+\delta\PSP(\RV) \label{eq:LipSha2}\\
\delta\PSP(\RV) & =  \iint\!\!d\RV' G_0\left(\RV-\RV';E\right)
 \hat{V}_2(\RV')\PSP(\RV') \nonumber \\
 & =\int\!\!dx'\left[2\Omega\cos(Kx')-\kappa\right]
 \left[G_0\left(x-x',y-\frac{a}{2};E\right) % & &  \nonumber \\
   +  pG_0\left(x-x',y+\frac{a}{2};E\right)\right]\PSP(x',\frac{a}{2}), \label{eq:LipShb2}
\end{align} 
where the factor $p$ is 1 for even and $-1$ for odd functions. The unperturbed function
$\PSP_0(y)$ in Eq.~(\ref{eq:LipSha2}) may be an even $\Psi_0^+(y)=\cos(k_yy)$ or an odd
$\Psi_0^-(y)=\sin(k_yy)$ eigenfunction for $k_x=0$ of the free motion Hamiltonian
$\hat{H_0}=(\hat{p}_{x}^2+\hat{p}_{y}^2)/2$. The corresponding Green's function $G_0$ is
given by Eq.~(\ref{eq:G02D}). The integration over $x'$ in  Eq.~(\ref{eq:LipShb2})
and over $\QV$ in Eq.~(\ref{eq:G02D}) reduce  the integral equation~(\ref{eq:LipSha2})
to the algebraic equation 
\begin{align}
& \PSP(\RV) =\sum_g\PPG(y)\exp(igx)=\PSP_0(y)\,+ \nonumber \\
 & \sum_g\exp(igx)
 \left[\Omega\left(\phi^p_{g-K}
 +\phi^p_{g+K}\right)-\kappa\PPG\right]  
 \left[\G1D\left(y-\frac{a}{2},E-\frac{g^2}{2}\right)
 +p\G1D\left(y+\frac{a}{2},E-\frac{g^2}{2}\right)\right],
 \label{eq:Laue2D2}
\end{align} 
where $\PPG\equiv\PPG(a/2)=p\PPG(-a/2)$, and $\G1D(y-y';E)$ is the free-electron
Green's function in 1D~\cite{Economou}, see Eq.~(\ref{eq:g01D}).
Similar to the case of a single wire, the sum over $g$ in the asymptotic
regions $y\rightarrow\pm\infty$ contains only finite number of terms with $g^2/2<E$.
In particular, for $k_y<K$, i.e., for the energies below $\ESB=K^2/2$, it contains only
the central beam, $g=0$:
\begin{equation}
\PPZ(y)=\PSP_0(y)\,+\left[\G1D\left(y+\frac{a}{2},E\right)
+p\G1D\left(y-\frac{a}{2},E\right)\right]
\left[\Omega\left(\phi^p_{-K}+\phi^p_K\right)-\kappa\PPZ\right]. \label{eq:cbeam2}
\end{equation}
\end{widetext}

As in the single-wire case, equating
the coefficients of the Fourier harmonics $\exp(igx)$ in Eq.~(\ref{eq:Laue2D2}) for $y=0$
leads to the recurrence relation for $\PPG$
\begin{flalign}
& \PPG(1\!+\!\kappa F^p_g)\!=\!
\delta_{g0}\PSP_0\left(\frac{a}{2}\right)
\!+\!\Omega F^p_g\left(\phi^p_{g-K}\!+\phi^p_{g+K}\right), & & \label{eq:recfig2}\\
& F^p_g \equiv \frac{1}{\zeta^p_g} \equiv  \G1D\left(0,E-\frac{g^2}{2}\right)
+p\G1D\left(a,E-\frac{g^2}{2}\right). & &  \nonumber 
\end{flalign}
The solution of the recurrence is similar to the relations (\ref{eq:recfig}), 
see Appendix \ref{sec:Calc-of-coefficients}, and it yields the perturbed solutions
for the even and odd standing waves.

Let us now consider a normally incident wave, for which the scattering function
is the combination
\begin{align}
\Psi(\RV) & =\Psi^+(\RV)+i\Psi^-(\RV)\label{eq:Psisum}\\
 & =\exp(iky)+\delta\Psi^+(\RV)+i\delta\Psi^-(\RV)\nonumber 
\end{align}

The wave function for the central beam is 
\begin{align}
\Psi_c(\RV) & =\phi^+_0(y)+i\phi^-_0(y) \label{eq:F0xy2}\\
 & =\begin{cases}
\exp(iky)+r_2\exp(-iky) & y\rightarrow-\infty\\
t_2\exp(iky) & y\rightarrow+\infty
\end{cases},
\end{align}
where $r_2$ and $t_2$ are the reflection and transmission amplitudes
\begin{align}
r_2 & =-\frac{2i}{k}\left\{ \cos\frac{ka}{2}\tilde{\phi}^+
-\sin\frac{ka}{2}\tilde{\phi}^-\right\} ,\label{eq:r2}\\
t_2 & =1-\frac{2i}{k}\left\{ \cos\frac{ka}{2}\tilde{\phi}^+
+\sin\frac{ka}{2}\tilde{\phi}^-\right\} ,\label{eq:t2}\\
\tilde{\phi}^p & \equiv 2\Omega\phi^p_K-\kappa\PPZ.\label{eq:fitild}
\end{align}
The coefficients $\phi^p_K$ and $\PPZ$ can be expressed as the continued
fraction of the same form as Eqs.~(\ref{eq:fiKcfT}) and~(\ref{eq:fi0cfT}),
respectively, with the functions $\zeta_g$ and $\theta_2$ replaced by
$\zeta_g^p$ and $\theta_2^p$. Substituting the resulting expressions for
$\phi^p_K$ and $\PPZ$ into Eq.~(\ref{eq:fitild}) we obtain after some algebra
\begin{align}
\tilde{\phi}^p  & =\frac{\PSP_0\left(\frac{a}{2}\right)
 \left[2\Omega^2-\kappa\left(\zeta^p_K+\kappa-\Omega^2\theta^p_2\right)
 \right]}{\left(\zeta^p_K+\kappa-\Omega^2\theta^p_2\right)
 \left(1+\kappa F^p_0\right)-2\Omega^2F^p_0},\label{eq:fitildfull} \\
F^p_0 & =-\frac{i}{k}[1+p\exp(ika)].\nonumber 
\end{align}
In order to simplify the expression~(\ref{eq:fitildfull}), we note
that for $\Omega=0$ it takes the form
\begin{equation}
\tilde{\phi}^p_0=-\frac{\kappa\PSP_0\left(\frac{a}{2}\right)}
{\left(1+\kappa F^p_0\right)}, \label{eq:fitild0}
\end{equation}
so we can write $\tilde{\phi}^p=\tilde{\phi}^p_0\tilde{\phi}^p_r$,
where 
\[
\tilde{\phi}^p_r \equiv\frac{\zeta^p_K
+\kappa-\Omega^2\left(\theta^p_2+2/\kappa\right)}{\zeta^p_K
+\kappa-\Omega^2\theta^p_2-\dfrac{2\Omega^2F^p_0}
{1+\kappa F^p_0}},
\]
which in the vicinity of the resonance energy $E^p_r\equiv (k^p_r)^2/2=\ESB -\EBD^p$
acquires the Fano shape
\begin{equation}
\tilde{\phi}^p_r \approx \frac{E-E^p_0}{E-E^p_{\textsc{p}} +i\Gamma^p_{\color{White}\textsc{p}}\!/2},
\label{eq:fitildF}
\end{equation}
where $\EBD^p=-(\kappa^p)^2/2$ is the energy of
the bound state of parity $p$ at $k_x=0$ of the uncorrugated double-wire.
The parameter $\kappa^p$ satisfies the equation
$\kappa^p a/[1+p\exp(-k^pa)] =\kappa a$, see Appendix~\ref{sec:BS}.
Its solution  for $\kappa a=1.5$ is illustrated in Fig.~\ref{f:eb2}.
Let us now calculate the parameters $E^p_0$, $\EPOL^p$, and $\Gamma^p$ in
Eq.~(\ref{eq:fitildF}). For $(E-E^p_r)/\kappa\ll 1$ we have
\[
\zeta^p_K +\kappa =\kappa-\frac{\sqrt{K^2-k^2}}{1+p\exp\left(-a\sqrt{K^2-k^2}\right)} 
\approx (E-E^p_r)\frac{\kappa C^p}
{(\kappa^p)^2},
\]
where we have introduced a dimensionless coefficient 
$C^p\equiv 1+p\kappa a\exp(-\kappa^p a)$.
This approximation gives for a given parity $p=\pm$ the transmission minimum at
\[
E^p_0 =E^p_r+\dfrac{(\Omega\kappa^p)^2\left(\theta^p_2+2/\kappa\right)}
{\kappa C^p},
\]
and the pole at the complex energy $\EPOL^p+i\Gamma^p/2$ with
$\EPOL^p =E^p_0-u^p$ and $\Gamma^p =2v^p$. 
Here $u^p$ and $v^p$ are the real and imaginary part of the quantity
$\dfrac{2(\Omega\kappa^p)^2}{\kappa^2C^p(1+\kappa F^p_0)}$.

We see that the reflection and transmission amplitudes for the double-wire
model, Eqs.~(\ref{eq:r2}) and (\ref{eq:t2}), contain a linear combination
of two Fano functions, $\tilde{\phi}^+$ and $\tilde{\phi}^-$.
Their interference gives a rather complicated transmissivity
and reflectivity functions $T_2(E)$ and $R_2(E)$, which depend in a complex 
way on the parameters of the potential $\hat{V}_2$, see Fig.~\ref{f:t2ls}.
Apart from that, one can see in Fig.~\ref{f:t2ls}(a) an additional transmission
resonance that is present both with and without corrugation,
see the blue dash-dotted curve. These are the $T=1$ resonances due to the 1D motion
that are the precursors of the energy bands of the crystal composed of the infinite
number of the wires. They are caused by the interference of the multiple reflections
between the wires.

\section{Conclusion}\label{sec:Concl}
We have demonstrated that the N-resonances in electron scattering on
atomically thin layers
occur due to the same mechanism as classical the Fano resonance, namely the
coupling between the bound and extended states. It is thus not surprising that the
transmissivity curve has the Fano shape. We have presented the 
mathematical proof and derived analytical expressions for the Fano parameters
for the specific case of corrugated wires, which establish a functional relation
between the complex energy of the resonance and the observed transmissivity. The \ai\
calculations for graphene~\cite{Nazarov2013} have turned out to be fully consistent with this
theory: We explained and quantitatively reproduced the results of Ref.~\cite{Nazarov2013},
in other words, we have traced the observed scattering resonance to the existence of a
strictly bound stationary state in an uncorrugated reference layer. 

Furthermore, we have shown that the scattering phase exhibits dramatic variation with energy
near the resonance, which results in a large Wigner time delay $\Delta\tau$. The \ai\
calculation of $\Delta\tau(E)$ for graphene has been found to closely follow the analytical
theory and agree with the resonance shape inferred from the scattering spectral function.

The theory has been generalized to several wires (layers in the 3D case) arranged in a
superstructure, whereby the bound states of the individual wires interact and the N-resonance
splits. In this case the scattering amplitude has the Fano shape near each resonance, and the
line-shape parameters depend in a rather complex way on the distance between the individual
scatterers.
   
\begin{acknowledgments}
  Fruitful discussions with V.U.~Nazarov are gratefully acknowledged. This work
  was supported by the Spanish Ministry of Science and Innovation (Projects
  No.~PID2022-139230NB-I00 and No.~PID2022-138750NB-C22) and by the National
  Academy of Sciences of Ukraine (Projects No.~III-2-22 and No.~III-4-23).
  D.E. and R.K. acknowledge support from
Volkswagen Foundation through the project "Synthesis, theoretical investigation and experimental study of emergent iron-based superconductors".  
\end{acknowledgments}

\appendix
%dummy comment 

\section{Fano Hamiltonian\label{sec:Fano-Hamiltonian}}

It is instructive to apply the Green's function approach to
the Fano Hamiltonian \cite{Fano1961}
\begin{align}
\hat{H}_{F} & =\hat{H}_0+\hat{V}_{F},\label{eq:HF}\\
\hat{H}_0 & =\left|\varphi\right\rangle E_{\varphi}\left\langle \varphi\right|
+\int\limits_{E_{\varphi}-D}^{E_{\varphi}+D}\left|\psi_{E}\right\rangle 
E\left\langle \psi_{E}\right|dE,\label{eq:H0F}\\
\hat{V}_{F} & =\int\limits_{E_{\varphi}-D}^{E_{\varphi}+D}
\left(\left|\varphi\right\rangle V_{E}^{*}\left\langle \psi_{E}\right|
+\left|\psi_{E}\right\rangle V_{E}\left\langle \varphi\right|\right)dE,\label{eq:VF}
\end{align}
where $\hat{H}_0$ is the unperturbed Hamiltonian, $|\varphi\rangle $ is 
the eigenfunction of its discrete energy level $E_{\varphi}$ that lies 
within the energy range
$E_{\varphi}-D < E <E_{\varphi}+D$, and $|\psi_{E}\rangle$ 
are the eigenfunctions of the continuous spectrum normalized as
$\langle \psi_{E^{\prime}}|\psi_{E}\rangle = \delta (E^{\prime}-E)$. 
The perturbation $\hat{V}$ couples these states. The eigenstates $|\varPhi\rangle$
of $\hat{H}_{F}$ of energy $E$ satisfy the Lippmann-Schwinger equation 
\begin{equation}
\left|\varPhi\right\rangle =\left|\psi_{E}\right\rangle +\hat G_0(E+i0)\hat{V}\left|\varPhi\right\rangle ,\label{eq:LSwF}
\end{equation}
where 
\[\hat G_0(\omega) =\left(\omega-\hat{H}_0\right)^{-1} 
=\frac{\left|\varphi\right\rangle \left\langle \varphi\right|}
{\omega-E_{\varphi}}+\int\limits_{E_{\varphi}-D}^{E_{\varphi}+D}
\frac{\left|\psi_{E}\right\rangle \left\langle \psi_{E}\right|}{\omega-E}dE
\]
is the resolvent of the unperturbed Hamiltonian, and we use the 
standard notation 
\[
f(E+i0)=\lim_{s\rightarrow0+}f(E+is).
\]
Substituting 
\[
\hat G_0(\omega)\hat{V}=\int\limits_{E_{\varphi}-D}^{E_{\varphi}+D}\left(\frac{\left|\varphi\right\rangle V_{E}^{*}\left\langle \psi_{E}\right|}{\omega-E_{\varphi}}+\frac{\left|\psi_{E}\right\rangle V_{E}\left\langle \varphi\right|}{\omega-E}\right)dE
\]
into Eq.(\ref{eq:LSwF}) yields
\begin{align}
\left|\varPhi\right\rangle & =\left|\psi_{E}\right\rangle 
 +\int\limits_{E_{\varphi}-D}^{E_{\varphi}+D}
\left(\frac{\left|\varphi\right\rangle V_{E^{\prime}}^{*}
\left\langle \psi_{E^{\prime}}|\varPhi\right\rangle }{E+i0-E_{\varphi}}\right.    \nonumber \\
& +\left.\frac{\left|\psi_{E^{\prime}}
\right\rangle V_{E^{\prime}}\left\langle \varphi|\varPhi\right\rangle }{E+i0-E^{\prime}}\right)dE^{\prime}.
\label{eq:LSwF1}
\end{align}
By multiplying Eq.~(\ref{eq:LSwF1}) from the left by $\left\langle \varphi\right|$
we obtain 
\begin{equation}
\left\langle \varphi|\varPhi\right\rangle =\int\limits_{E_{\varphi}-D}^{E_{\varphi}+D}\frac{V_{E^{\prime}}^{*}\left\langle \psi_{E^{\prime}}|\varPhi\right\rangle }{E+i0-E_{\varphi}}dE^{\prime},\label{eq:fiFi}
\end{equation}
and the multiplication by $\left\langle \psi_{E^{\prime\prime}}\right|$ gives
\begin{equation}
\left\langle \psi_{E^{\prime\prime}}|\varPhi\right\rangle =\delta\left(E-E^{\prime\prime}\right)+\frac{V_{E^{\prime\prime}}\left\langle \varphi|\varPhi\right\rangle }{E+i0-E^{\prime\prime}}.\label{eq:psiFi}
\end{equation}\\
Upon substituting Eq.~(\ref{eq:psiFi}) into Eq.~(\ref{eq:fiFi}) we obtain 
\begin{align}
\left\langle \varphi|\varPhi\right\rangle  & =\frac{V_{E}^{*}}{E+i0-E_{\varphi}-F(E)},\label{eq:fiFi1}\\
F(E) & \equiv\int\limits_{E_{\varphi}-D}^{E_{\varphi}+D}\frac{\left|V_{E^{\prime}}\right|^2}{E+i0-E^{\prime}}dE^{\prime}.\nonumber 
\end{align}
This gives 
\begin{align}
\left\langle \psi_{E^{\prime\prime}}|\varPhi\right\rangle & 
 =\delta\left(E-E^{\prime\prime}\right)     \nonumber \\
 &+\frac{V_{E^{\prime\prime}}}{E+i0-E^{\prime\prime}}\frac{V_{E}^{*}}{E+i0-E_{\varphi}-F(E)}.
 \label{eq:psiFi1}
\end{align}
Finally, we substitute Eqs.~(\ref{eq:fiFi1}) and (\ref{eq:psiFi1})
into Eq.~(\ref{eq:LSwF1}) and obtain 
\begin{widetext}
\begin{equation}
\left|\varPhi\right\rangle =\left|\psi_{E}\right\rangle 
+\frac{V_{E}^{*}}{E+i0-E_{\varphi}-F(E)}\left[\left|\varphi\right\rangle
 +\int\limits_{E_{\varphi}-D}^{E_{\varphi}+D}\frac{V_{E^{\prime}}}{E+i0-E^{\prime}}
 \left|\psi_{E^{\prime}}\right\rangle dE^{\prime}\right].\label{eq:FipmF}
\end{equation}  \end{widetext}
It is easy to verify that $\hat{H}_{F}\left|\varPhi\right\rangle =E\left|\varPhi\right\rangle $.

Let us consider the $r\rightarrow\infty$ asymptotic of $\left\langle r|\varPhi\right\rangle $.
The eigenfunction of the discrete energy level is localized, and $\left\langle r|\varphi\right\rangle \rightarrow0$.
For the eigenstates that belong to the continuous spectrum, one have
$\left\langle r|\psi_{E^{\prime}}\right\rangle \propto\sin[k(E')r+\delta]$,
where $k(E')=\sqrt{2E'}$, $\delta$ is a phase shift \cite{Fano1961,Landafshitz1981quantum}.
It is reasonable to assume that the matrix element between the localized
and extended functions $V_{E^{\prime}}$ decreases rapidly as a function
of $|E'-E_{\varphi}|$. Then, passing to the integration over $k$,
we may extend the limits of integration to $\pm\infty$ and apply
the residue theorem for the calculation of the integral \cite{Fano1961}
\[
\int\limits_{-\infty}^{+\infty}\frac{V(k^2/2)\sin[k(E')+\delta]}{E\pm i0-k^2/2}kdk=-\pi V_{E}\cos[k(E)r+\delta].
\]
Then 
\begin{align*}
\left\langle r|\varPhi\right\rangle  & \propto\sin[k(E)r+\delta]-\frac{\pi|V_{E}|^2\cos[k(E)r+\delta]}{E+i0-E_{\varphi}-F(E)}\\
 & \propto\sin[k(E)r+\delta+\Delta_{r}],
\end{align*}
where the resonance part of the phase shift is (cf. Eq. (15) of Ref.
\cite{Fano1961}) 
\begin{align*}
\Delta_{r} & =-\arctan\frac{\pi|V_{E}|^2}{E-E_{\varphi}-F(E)}-\arctan\frac{\Gamma}{2(E-\EPOL)},
\end{align*}
where $\Gamma\equiv2\pi|V_{E}|^2$ and $\EPOL\equiv E_{\varphi}+F(E)$.
The sharp variation of the phase shift as $E$ passes through the
resonance energy $\EPOL$ leads to a large Wigner time delay in a
wave packet scattering, see Eq. (\ref{eq:dtaurapp}).

\section{Calculation of coefficients $\phi_{nK}$ \label{sec:Calc-of-coefficients}}

Let us find the solution of the recurrences (\ref{eq:recfig}) and (\ref{eq:recfig2}).
As mentioned in the main text, it follows from the symmetry of the scattering
function that the coefficients of Laue expansions, Eqs.~(\ref{eq:Laue}) and
(\ref{eq:Laue2}) obey the relation $\phi_g=\phi_{-g}$. Here we drop the superscript
$p$ in the recurrence terms of Eq.~(\ref{eq:recfig2}); it is easily restored in the
final formulas.
We rewrite the equations (\ref{eq:recfig}) and (\ref{eq:recfig2}) as 
\begin{equation}
\zeta_g+\kappa-\Omega\frac{\phi_{g+K}}{\phi_g}=\frac{\zeta_0b_0}{\phi_0}\delta_{g0}
+\frac{\Omega\phi_{g-K}}{\phi_g},\label{eq:irec}
\end{equation}
where $b_0=1$ in Eq.~(\ref{eq:recfig}) and $b_0=\PSP_0\left(\frac{a}{2}\right)$ in
Eq.~(\ref{eq:recfig2}). For $g=0$, Eq.~(\ref{eq:irec})  gives
\begin{equation}
\phi_0=\dfrac{b_0\zeta_0}{\zeta_0
+\kappa-2\Omega\dfrac{\phi_{K}}{\phi_0}},\label{eq:fiK2fi0}
\end{equation}
and for $g>0$ the recurrence is 
\begin{equation}
\dfrac{\phi_g}{\phi_{g-K}}=\dfrac{\Omega}{\zeta_g
+\kappa-\Omega\dfrac{\phi_{g+K}}{\phi_g}}.\label{eq:figpK2fig}
\end{equation}
Substituting these expressions for consequent $g$ into Eq.~(\ref{eq:fiK2fi0}),
we obtain the solution for $\phi_0$ in the form of continued fraction 
\begin{align}
\phi_0 & =\cfrac{b_0\zeta_0}{\zeta_0+\kappa-\cfrac{2\Omega^2}{\zeta_K
+\kappa-\Omega\cfrac{\phi_{2K}}{\phi_{K}}}} 
=\cfrac{b_0\zeta_0}{\zeta_0+\kappa-2\Omega^2\theta_1}, \label{eq:fi0cf}
\end{align}
where the function $\theta_1(E)$ is 
\begin{equation}
\theta_1  =\cfrac{1}{\zeta_K+\kappa-\Omega^2\theta_2},\label{eq:theta1}
\end{equation}
and for $n>1$ the recurrence reads
\begin{equation}
\theta_{n} =\cfrac{1}{\zeta_{nK}+\kappa-\Omega^2\theta_{n+1}}.\label{eq:thetan}
\end{equation} 
Thus, $\theta_1(E)$ is the continued fraction
\begin{equation}
\theta_1 =\cfrac{1}{\zeta_K+\kappa
-\cfrac{\Omega^2}{\zeta_{2K}+\kappa
-\cfrac{\Omega^2}{\zeta_{3K}+\kappa-\cfrac{\Omega^2}{\ddots}}}}. \label{eq:theta1cf}
\end{equation} 
which can be calculated with any desired accuracy. We may truncate
the chain by setting $\phi_{n_0K} = 0$, i.e., $\theta_{n0}= 0$, at different $n_0$, for example:

$n_0=3$: setting $\phi_{3K}= 0$ and $\theta_{3}= 0$ yields
\begin{equation}
\theta_{1,3}\approx  
\cfrac{1}{\zeta_K+\kappa-\cfrac{\Omega^2}{\zeta_{2K}+\kappa}}; \label{eq:theta13}
\end{equation}

$n_0=4$: setting $\phi_{4K}=0$ and $\theta_{4}=0$ yields
\begin{equation}
\theta_{1,4}\approx
\cfrac{1}{\zeta_K+\kappa-\cfrac{\Omega^2}{\zeta_{2K}+\kappa
-\cfrac{\Omega^2}{\zeta_{3K}+\kappa}}}.\label{eq:theta14}
\end{equation}

Alternatively, $\theta_1$ can be calculated by the modified Lentz's 
algorithm \cite{Lentz1976,Thompson1986,NumRecipes}. Then the function
$f\equiv 1/\theta_1$ is calculated by iterations 
\begin{align}
f_j & =f_{j-1}\Delta _j, \quad \Delta _j =C_jD_j,  \nonumber \\ %%% \label{eq:fiter}\\
C_j & = b_j+\frac{a_j}{C_{j-1}},\quad 
D_j =\frac{1}{b_j+a_jD_{j-1}}, \quad j=1,2\dots , \nonumber
\end{align}
where $f_0=C_0=b_0$, $D_0=0$, $b_j=\zeta _{(j+1)K}+\kappa$, and $a_j=-\Omega ^2$.
The iterations stop when $|\Delta _j-1| \approx 0$. More details can be found in
Refs.~\onlinecite{Lentz1976,Thompson1986,NumRecipes}. It is the result of the iterative
procedure that we have referred to as numerically exact.

Having calculated $\theta_1$, we obtain $\phi_0$ from Eq.~(\ref{eq:fi0cf}).
Comparing Eqs.~(\ref{eq:fi0cf}) and~(\ref{eq:fiK2fi0}), we see that
$\phi_{K}=\Omega\phi_0\theta_1$.
The numerically exact transmissivity is depicted in Fig.~\ref{f:tr} by solid
line. The deviation of the approximations by Eq.~(\ref{eq:theta13}) or
(\ref{eq:theta14}) from the exact result is within the linewidth.

%%%%%%%%%%%%%%%%%%%%%%%%%%%%%%%%%%%%%%%%%%%%%%%%%%%%%%%%%%%%%% 
\begin{figure}
\centering \includegraphics[width=0.9\columnwidth]{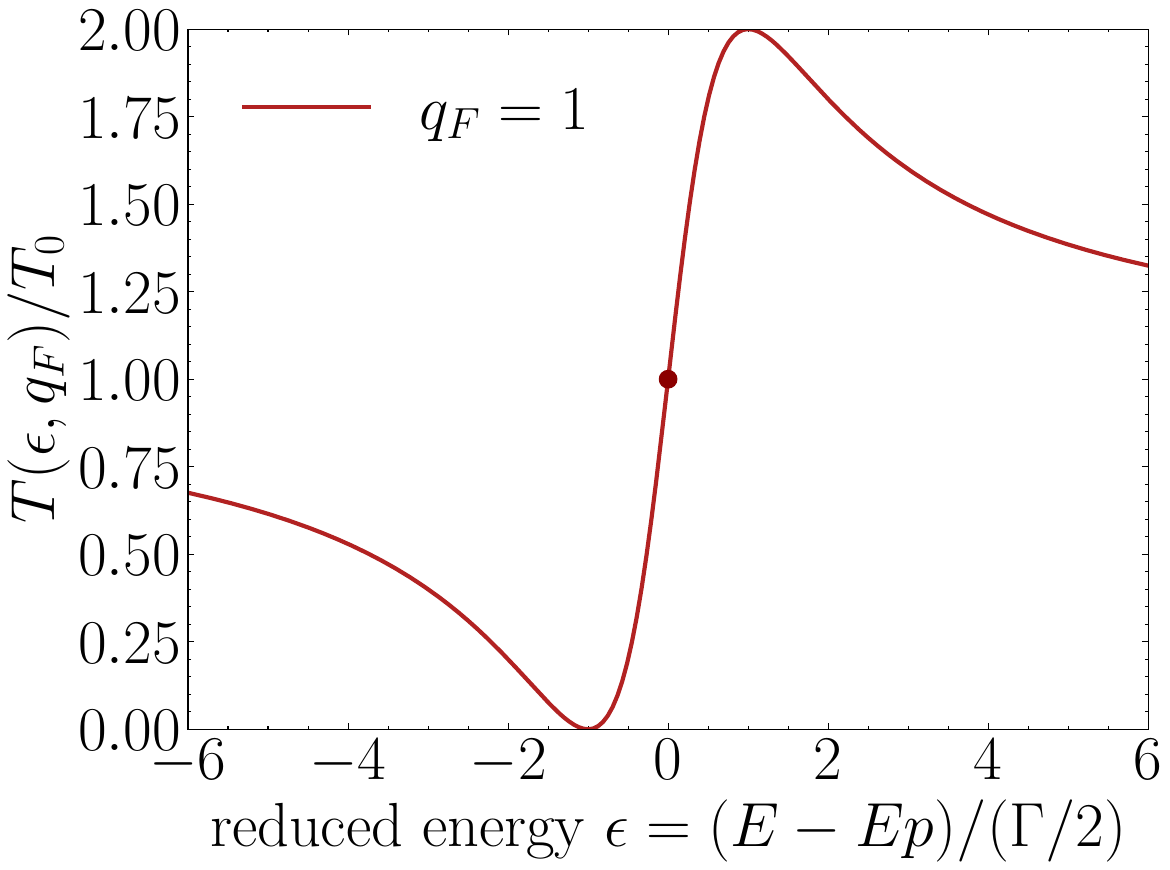} 
\caption{Reduced transmission probability % $T(\epsilon,\QF)/T_0$ 
near the Fano resonance. The circle indicates the reduced energy $\EPSD$ at which
$T(\EPSD)=T(\EPSX)/2$.}
\label{f:Fano} 
\end{figure}

%%%%%%%%%%%%%%%%%%%%%%%%%%%%%%%%%%%%%%%%%%%%%%%%%%%%%%%%%%%%%
%%%%%%%%%%%%%%%%%%%%%%%%%%%%%%%%%%%%%%%%%%%%%%%%%%%%%%%%%%%%%% 
\section{Analysis of Fano resonance curve}
\label{sec:expFano}

Here we show how to infer the real and imaginary parts of the pole, $\EPOL$
and $\Gamma/2$, from the energies of the minimum (zero) of the experimental
curve $E_0$, its maximum $\EMAX$, and the point $\EMID$, where
the transmissivity is half of its maximum.
Near a Fano resonance, the transmission amplitude has the form
given by Eq.~(\ref{eq:tFanoamp}). The transmissivity reads
\begin{align}
T(E) =|t|^2=T_0\frac{(E-E_0)^2}{(E-\EPOL)^2+(\Gamma/2)^2}.\label{eq:TFano1}
\end{align}
From Eq.~(\ref{eq:tFanoamp}) we see that the amplitude has a pole at 
$z=\EPOL-i\Gamma/2$.
Strictly speaking, the pole occurs in the analitic continuation of the
amplitude to the lower complex-energy half-plane of the unphysical sheet
of the Riemann surface (through the branch cut on the real axis for E > 0).

If one obtains the curve $T(E)$ from a (numerical) experiment, the
parameters $\EPOL$ and $\Gamma$ are not known. Below, we show how
they may be obtained from three characteristic features of the curve.
Namely, the minimum (zero) of the experimental curve $E_0$, its
maximum, $\EMAX$ and the point $\EMID$, where $T(\EMID)=T(\EMAX)/2$

The expression (\ref{eq:TFano1}) is usually written as a function
of the reduced variables $\QF$ (\ref{eq:qF}) and $\epsilon$: 
\begin{align}
T(\epsilon,\QF) & =T_0\frac{(\epsilon+\QF)^2}{\epsilon^2+1},\label{eq:Teq}\\
\epsilon & \equiv(E-\EPOL)/(\Gamma/2),\label{eq:eps}
\end{align}

Figure~\ref{f:Fano} shows the reduced transmission probability $T/T_0$ for $\QF=1$.
The curve is described by the three
characteristic points: the minimum $T(\epsilon_0,\QF)=0$ at $\epsilon_0=-\QF$, 
the maximum $T(\EPSX,\QF)=T_0(\QF^2+1)$ at $\EPSX=1/\QF$, and the midpoint 
$T(\EPSD,\QF)=T(\EPSX,\QF)/2$
at $\EPSD=(1-\QF)/(1+\QF)$. Using the definitions (\ref{eq:eps}) and (\ref{eq:qF}),
we obtain 
\begin{align}
\Delta_{m} & =\EMAX-E_0=\frac{\Gamma(\QF^2+1)}{2\QF},\label{eq:Dm2G}\\
\EPSD-\epsilon_0 & =\frac{\QF^2+1}{\QF+1},\nonumber \\
\Delta_0 & =\EMID-E_0=\frac{\Gamma(\QF^2+1)}{2(\QF+1)}\nonumber 
=\frac{\QF\Delta_{m}}{\QF+1},\nonumber 
\end{align}
where $\EMID$ is the energy such that $T(\EMID)=T(\EMAX)/2$, 
which is situated between the minimum at $E_0$ and maximum at $\EMAX$. In the last line,
we have used the relation between $\Delta_{m}$ and $\Gamma$, given by Eq.(\ref{eq:Dm2G}).
Thus, we may find the parameter $\QF$ from the energy
differences that we know from the experimental curve 
\begin{equation}
\QF=\frac{\Delta_0}{\Delta_{m}-\Delta_0}=\frac{\EMID-E_0}{\EMAX-\EMID},\label{eq:qfromdlt}
\end{equation}
which allows us to obtain the imaginary part of the pole position from Eq.~(\ref{eq:Dm2G}) 
\begin{equation}
\Gamma=\frac{2\QF\Delta_{m}}{\QF^2+1},\label{eq:Dm2Gq}
\end{equation}
and the real part 
\begin{equation}
\EPOL=E_0+\frac{\QF\Gamma}{2}.\label{eq:qG2Ep}
\end{equation}

\section{Time delay}
\label{sec:delay}

Here we give the explicit formulas for the $\dot{\eta_{r}}$ that
are valid in the energy range $0<E<K^2/2$ . In the Eq.~(\ref{eq:tfull}),
we denote 
\begin{align*}
\varepsilon & \equiv \zeta_K+\kappa,\\
\varepsilon_0 & \equiv \Omega^2\kappa\theta_2,\\
\VPOL & \equiv\varepsilon_0+\dfrac{2\Omega^2\kappa^2}{k^2+\kappa^2},\\
\gamma/2 & \equiv\dfrac{2\Omega^2\kappa k}{k^2+\kappa^2}.
\end{align*}
Then the resonance part of the scattering phase $\eta_{r}$, Eq. (\ref{eq:tgetar}),
takes the form 
\[
\eta_{r}=-\arctan\frac{\gamma}{2(\varepsilon-\VPOL)}.
\]
Its energy derivative is 
\[
\dot{\eta}_{r}=-\frac{\dot{\gamma}(\varepsilon-\VPOL)-\gamma(\dot{\varepsilon}-\VPOD)}{2\left[(\varepsilon-\VPOL)^2+(\gamma/2)^2\right]},
\]
where 
\begin{align*}
\dot{\varepsilon} & =\frac{\kappa}{\sqrt{K^2-k^2}},\\
\VPOD & =\Omega^2\kappa\dot{\theta}_2-\dfrac{4\Omega^2\kappa^2}{(\kappa^2+k^2)^2},\\
\dot{\gamma} & =\dfrac{4\Omega^2\kappa(\kappa^2-k^2)}{k(\kappa^2+k^2)^2},\\
\dot{\theta}_2 & \approx-\dfrac{1}{\sqrt{4K^2-k^2}\left(\kappa-\sqrt{4K^2-k^2}\right){}^2}.
\end{align*}
In the last line we have used the $n_0=3$ approximation for $\theta_2$,
see Eq.~(\ref{eq:theta13}), 
\begin{align*}
  \theta_2 \approx\frac{1}{\kappa+\zeta_{2K}} 
  =\frac{1}{\kappa-\sqrt{4K^2-k^2}}.
\end{align*}
In the $n_0=4$ approximation, see Eq.~(\ref{eq:theta14}), we obtain
\begin{align*}
& \theta_{2,4} \approx\dfrac{1}{\zeta_{2K}+\kappa-\dfrac{\Omega^2}{\zeta_{3K}+\kappa}},\\
& \dot{\theta}_{2,4} \approx \\
& -\theta_{2,4}^2\left[\dfrac{1}{\sqrt{4K^2-k^2}}
+\dfrac{\Omega^2}{\sqrt{9K^2-k^2}\left(\kappa-\sqrt{9K^2-k^2}\right)^2}\right].
\end{align*}

\section{Bound states}
\label{sec:BS}

For two wires, for $\Omega=0$ the motion along $\YDIR$ is described by a 1D Hamiltonian 
\begin{equation}
\hat{H}_{y,2}=\frac{\hat{p}_{y}^2}{2}-\kappa\left[\delta\left(y-\frac{a}{2}\right)
+\delta\left(y+\frac{a}{2}\right)\right]\label{eq:Hy2}
\end{equation}
The bound states, having energy $\EBD^p<0$ are easily found from the
Lippmann-Schwinger equation
\begin{align}
\varphi_{b}(y) & =-\kappa\int\!\!dy'\G1D\left(y-y';\EBD^p\right)\times \nonumber \\
& \times\left[\delta\left(y'-\frac{a}{2}\right)+\delta\left(y'+\frac{a}{2}\right)\right]\varphi_{b}(y'),
\label{eq:LipShaBS}
\end{align}
where $\G1D\left(y-y';E\right)$ is the free-electron Green's function
in 1D~\cite{Economou}, Eq.~(\ref{eq:g01D}). Equations~(\ref{eq:LipShaBS}) can be solved
separately for even and odd functions. After integration and substitution of the expression
(\ref{eq:g01D}) for the Green's function, Eq.~(\ref{eq:LipShaBS}) reduces to 
\begin{equation}
\varphi^p_\textsc{b}(y)=\frac{\kappa}{\kappa^p}\left[e^{-\kappa^p|y-a/2|}
+pe^{-\kappa^p|y+a/2|}\right]\varphi^p_\textsc{b}\left(\frac{a}{2}\right),\label{eq:LibSaBSa}
\end{equation}
where $\kappa^p=\sqrt{2|\EBD^p|}$. From Eq.~(\ref{eq:LibSaBSa}) we obtain for $y=a/2$
\begin{align}
\kappa^p & =\kappa\left(1+pe^{-\kappa^pa}\right)\label{eq:kab}\\
 & \approx\kappa\left(1+pe^{-\kappa a}\right).\label{eq:kabappr}
\end{align}
The last approximate equality holds for $\exp(-\kappa a)\ll 1$, i.e., in the
tight-binding regime. Generally, Eq.~(\ref{eq:kab}) should be solved numerically.
The solution is visualized in Fig.~\ref{f:eb2}. We see that the solution for the
even bound state $p = +$ exists for all values of the potential strength $\kappa$
and the interwire distance $a$, whereas the odd bound states, $p = -$,
exist only for sufficiently large values of the product $\kappa a > 1$.
Then, in terms of $\kappa^p$ the binding energy is
\begin{equation}
\EBD^p=-\frac{(\kappa^p)^2}{2}. \nonumber %%% \label{eq:Ebalf}
\end{equation}

\end{document}